\documentclass[zpreprint,zbstnp]{zeus_paper}
\usepackage[english]{babel}
\chardef\usc=95
\chardef\til=126
\catcode`\@=11 
\DeclareRobustCommand\xdotspace{\futurelet\@let@token\@xdotspace}
\def\@xdotspace{%
  \ifx\@let@token.\else
  \ifx\@let@token\bgroup.\else
  \ifx\@let@token\egroup.\else
  \ifx\@let@token\/.\else
  \ifx\@let@token\ .\else
  \ifx\@let@token~.\else
  \ifx\@let@token!.\else
  \ifx\@let@token,.\else
  \ifx\@let@token:.\else
  \ifx\@let@token;.\else
  \ifx\@let@token?.\else
  \ifx\@let@token/.\else
  \ifx\@let@token'.\else
  \ifx\@let@token).\else
  \ifx\@let@token-.\else
  \ifx\@let@token\@xobeysp.\else
  \ifx\@let@token\space.\else
  \ifx\@let@token\@sptoken.\else
   .\space
   \fi\fi\fi\fi\fi\fi\fi\fi\fi\fi\fi\fi\fi\fi\fi\fi\fi\fi}
\catcode`\@=12 

\newcommand{\stru}[2]{%
   \relax\ifmmode\hbox{\vrule height#1 depth#2 width0pt}%
   \else\vrule height#1 depth#2 width0pt\fi}

\newcommand{\Ronum}[1]{\uppercase\expandafter{\romannumeral#1}}
\newcommand{\ronum}[1]{\expandafter{\romannumeral#1}}
\DeclareRobustCommand{\LaTeXZ}{%
  \LaTeX\kern-.05em4\kern-.1em
  {\raisebox{-0.2ex}{$\scriptstyle\text{ZEUS}$}}\xspace}

\DeclareMathAlphabet{\mathbf}{OT1}{cmr}{bx}{sl}
\newcommand{\eVdist}{\kern-0.06667em}

\newcommand{\slashfrac}[2]{%
  \raisebox{0.5ex}{\ensuremath #1}\kern-0.12em/\kern-0.08em
  \raisebox{-.8ex}{\ensuremath #2}}

\newcommand{\sqr}[3]{%
    {\vcenter{\hrule height.#3ex\hbox{\vrule width.#2ex height#1ex
     \kern#1ex\vrule width.#3ex}\hrule height.#2ex}}}

\catcode`\@=11 
\newcommand{\parenbar}{\mathpalette\p@renb@r}
\def\p@renb@r#1#2{\vbox{%
  \ifx#1\scriptscriptstyle \dimen@.7em\dimen@ii.2em\else
  \ifx#1\scriptstyle \dimen@.8em\dimen@ii.25em\else
  \dimen@1em\dimen@ii.4em\fi\fi \offinterlineskip
  \ialign{\hfill##\hfill\cr
    \vbox{\hrule width\dimen@ii}\cr
    \noalign{\vskip-.3ex}%
    \hbox to\dimen@{$\mathchar300\hfil\mathchar301$}\cr
    \noalign{\vskip-.3ex}%
    $#1#2$\cr}}}
\catcode`\@=12 

\newcommand{\IP}{{\rm I$\kern-0.01667em$P}\xspace}

\mathchardef\qsm=63
\mathchardef\pls=43
\mathchardef\mns=512
\mathchardef\plm=518
\mathchardef\eql=61
\mathchardef\smallleft=300
\mathchardef\smallright=301
\mathchardef\les=316
\mathchardef\gre=318
\mathchardef\leq=532
\mathchardef\grq=533

\catcode`\@=11 
\newcounter{pict@width}
\newcounter{pict@height}
\newlength{\pict@scale}
\newcommand{\psfigadd}[4]{%
\setcounter{pict@width}{1*\ratio{#2+\pict@scale/2}{\pict@scale}}
\setcounter{pict@height}{1*\ratio{#3+\pict@scale/2}{\pict@scale}}
\setlength{\unitlength}{\pict@scale}
\hbox to #2{\hspace{-\fill}\begin{picture}(\thepict@width,\thepict@height)
\put(0,0){\psfig{figure=#1,width=#2,height=#3,clip=}}
\SetScale{0.283466457}
\SetWidth{1.763889}
{#4}
\end{picture}}
}
\newcounter{pict@widthfst}
\newcounter{pict@widthscd}
\newcounter{pict@widthtot}
\newcommand{\psfigaddtwo}[7]{%
\setcounter{pict@widthfst}{1*\ratio{#2+\pict@scale/2}{\pict@scale}}
\setcounter{pict@widthscd}{1*\ratio{#2+#4+\pict@scale/2}{\pict@scale}}
\setcounter{pict@widthtot}{1*\ratio{#2+#4+#6+\pict@scale/2}{\pict@scale}}
\setcounter{pict@height}{1*\ratio{#3+\pict@scale/2}{\pict@scale}}
\setlength{\unitlength}{\pict@scale}
\hbox{\hspace{-\fill}\begin{picture}(\thepict@widthtot,\thepict@height)
\put(0,0){\psfig{figure=#1,width=#2,height=#3,clip=}}
\put(\thepict@widthscd,0){\psfig{figure=#5,width=#6,height=#3,clip=}}
\SetScale{0.283466457}
\SetWidth{1.763889}
{#7}
\end{picture}}
}
\newcommand{\psfigror}[4]{%
\setcounter{pict@width}{1*\ratio{#2+\pict@scale/2}{\pict@scale}}
\setcounter{pict@height}{1*\ratio{#3+\pict@scale/2}{\pict@scale}}
\setlength{\unitlength}{\pict@scale}
\hbox{\begin{picture}(\thepict@width,\thepict@height)
\put(0,\thepict@height){\psfig{figure=#1,width=#3,height=#2,clip=,angle=270}}
\SetScale{0.283466457}
\SetWidth{1.763889}
{#4}
\end{picture}}
}
\newcommand{\psfigrol}[4]{%
\setcounter{pict@width}{1*\ratio{#2+\pict@scale/2}{\pict@scale}}
\setcounter{pict@height}{1*\ratio{#3+\pict@scale/2}{\pict@scale}}
\setlength{\unitlength}{\pict@scale}
\hbox{\begin{picture}(\thepict@width,\thepict@height)
\put(0,0){\psfig{figure=#1,width=#3,height=#2,clip=,angle=90}}
\SetScale{0.283466457}
\SetWidth{1.763889}
{#4}
\end{picture}}
}
\catcode`\@=12 
\newlength\listtextwidth

\catcode`\@=11 
\newlength{\@tabfninsert}
\newlength{\@tabfnwidth}
\newcommand{\tabfootnote}[2]{%
  \setlength{\@tabfninsert}{0.8em}
  \setlength{\@tabfnwidth}{\textwidth}
  \addtolength{\@tabfnwidth}{-\@tabfninsert}
  \addtolength{\@tabfnwidth}{-0.4em}
  \noindent\makebox[\@tabfninsert][r]{\footnotesize$^{#1}$\hfil}\hfill%
  \parbox[t]{\@tabfnwidth}{\footnotesize #2\hfill}}
\catcode`\@=12 

\newcommand {\pom} {I\!\!P}
\newcommand {\reg} {I\!\!R}
\newcommand {\xpom} {x_{\pom}}

\def\citeCAL{{\cite{%
nim:a309:77,*nim:a309:101,*nim:a321:356,*nim:a336:23%
}}\xspace}

\def\citediff0{{\cite{%
zeusdiff0,*h1diff0%
}}\xspace}

\def\bpc1{{\cite{%
bpc01,*bpc02,*bpc03%
}}\xspace}

\def\lsim{\mathrel{\rlap{\lower4pt\hbox{\hskip1pt$\sim$}}
    \raise1pt\hbox{$<$}}}                
\def\gsim{\mathrel{\rlap{\lower4pt\hbox{\hskip1pt$\sim$}}
    \raise1pt\hbox{$>$}}}                

\begin{document}
\prepnum{{DESY--08--175}}
\title{
Deep inelastic scattering with leading protons or large rapidity gaps 
at HERA
}
\author{ZEUS Collaboration}
\date{December 2008}
\abstract{ The dissociation of virtual photons, $\gamma^{\star} p \to X 
p$, in events with a large rapidity gap between $X$ and the outgoing 
proton, as well as in events in which the leading proton was directly 
measured, has been studied with the ZEUS detector at HERA. The data cover 
photon virtualities $Q^2>2$~GeV$^2$ and $\gamma^{\star} p$ centre-of-mass 
energies 
$40<W<240$ GeV, with $M_X>2$ GeV, where $M_X$ is the mass of the hadronic 
final state, $X$. Leading protons were detected in the ZEUS leading proton 
spectrometer. The cross section is presented as a function of $t$, the 
squared four-momentum transfer at the proton vertex and $\Phi$, the 
azimuthal angle between the positron scattering plane and the proton 
scattering plane. It is also shown as a function of $Q^2$ and $\xpom$, the 
fraction of the proton's momentum carried by the diffractive exchange, as 
well as $\beta$, the Bjorken variable defined with respect to the 
diffractive exchange.}
\makezeustitle
\def\3{\ss}
\pagenumbering{Roman}
%
%
%
%
%
%
%
\pagenumbering{Roman}                                                                              
\begin{center}                                                                                     
{                      \Large  The ZEUS Collaboration              }                               
\end{center}                                                                                       
  S.~Chekanov,                                                                                     
  M.~Derrick,                                                                                      
  S.~Magill,                                                                                       
  B.~Musgrave,                                                                                     
  D.~Nicholass$^{   1}$,                                                                           
  \mbox{J.~Repond},                                                                                
  R.~Yoshida\\                                                                                     
 {\it Argonne National Laboratory, Argonne, Illinois 60439-4815, USA}~$^{n}$                       
\par \filbreak                                                                                     
  M.C.K.~Mattingly \\                                                                              
 {\it Andrews University, Berrien Springs, Michigan 49104-0380, USA}                               
\par \filbreak                                                                                     
  P.~Antonioli,                                                                                    
  G.~Bari,                                                                                         
  L.~Bellagamba,                                                                                   
  D.~Boscherini,                                                                                   
  A.~Bruni,                                                                                        
  G.~Bruni,                                                                                        
  G.~Cara~Romeo                                                                                    
  F.~Cindolo,                                                                                      
  M.~Corradi,                                                                                      
\mbox{G.~Iacobucci},                                                                               
  A.~Margotti,                                                                                     
  T.~Massam,                                                                                       
  R.~Nania,                                                                                        
  A.~Polini\\                                                                                      
  {\it INFN Bologna, Bologna, Italy}~$^{e}$
% ...
\newpage                                                                                           
\enlargethispage{5cm}                                                                              
$^{\    1}$ also affiliated with University College London,                                        
United Kingdom\\                                                                                   
$^{\    2}$ now at University of Salerno, Italy \\                                                 
% ...
$^{\    5}$ supported by the research grant no. 1 P03B 04529 (2005-2008) \\                        
% ...
%
\newpage   
                                                           %
                                                           %
\begin{tabular}[h]{rp{14cm}}                                                                       
$^{a}$ &  supported by the Natural Sciences and Engineering Research Council of Canada (NSERC) \\  
$^{b}$ &  supported by the German Federal Ministry for Education and Research (BMBF), under        
          contract numbers 05 HZ6PDA, 05 HZ6GUA, 05 HZ6VFA and 05 HZ4KHA\\                         
$^{c}$ &  supported in part by the MINERVA Gesellschaft f\"ur Forschung GmbH, the Israel Science   
          Foundation (grant no. 293/02-11.2) and the U.S.-Israel Binational Science Foundation \\  
$^{d}$ &  supported by the Israel Science Foundation\\                                             
$^{e}$ &  supported by the Italian National Institute for Nuclear Physics (INFN) \\                
$^{f}$ &  supported by the Japanese Ministry of Education, Culture, Sports, Science and Technology 
          (MEXT) and its grants for Scientific Research\\                                          
$^{g}$ &  supported by the Korean Ministry of Education and Korea Science and Engineering          
          Foundation\\                                                                             
$^{h}$ &  supported by the Netherlands Foundation for Research on Matter (FOM)\\                   
$^{i}$ &  supported by the Polish State Committee for Scientific Research, project no.             
          DESY/256/2006 - 154/DES/2006/03\\                                                        
$^{j}$ &  partially supported by the German Federal Ministry for Education and Research (BMBF)\\   
$^{k}$ &  supported by RF Presidential grant N 1456.2008.2 for the leading                         
          scientific schools and by the Russian Ministry of Education and Science through its      
          grant for Scientific Research on High Energy Physics\\                                   
$^{l}$ &  supported by the Spanish Ministry of Education and Science through funds provided by     
          CICYT\\                                                                                  
$^{m}$ &  supported by the Science and Technology Facilities Council, UK\\                         
$^{n}$ &  supported by the US Department of Energy\\                                               
$^{o}$ &  supported by the US National Science Foundation. Any opinion,                            
findings and conclusions or recommendations expressed in this material                             
are those of the authors and do not necessarily reflect the views of the                           
National Science Foundation.\\                                                                     
$^{p}$ &  supported by the Polish Ministry of Science and Higher Education                         
as a scientific project (2006-2008)\\                                                              
$^{q}$ &  supported by FNRS and its associated funds (IISN and FRIA) and by an Inter-University    
          Attraction Poles Programme subsidised by the Belgian Federal Science Policy Office\\     
$^{r}$ &  supported by an FRGS grant from the Malaysian government\\                               
\end{tabular}                                                                                      
                                                           %
                                                           %

%
%
%
\newpage
%
%
%
\pagenumbering{arabic} 
\pagestyle{plain}

\section{Introduction}
\label{sec-int}

In diffractive hadron-hadron or photon-hadron collisions, the 
initial-state particles undergo a ``peripheral" collision, in which they 
either stay intact (elastic scattering), or dissociate into low-mass 
states (diffractive dissociation). The scattered hadron (or the low-mass 
state in the dissociative case) has energy equal, to within a few per 
cent, to that of the incoming hadron, and very small transverse momentum. 
Such interactions can be described in the framework of Regge 
phenomenology, where they are ascribed to the exchange of a trajectory 
with the vacuum quantum numbers, the Pomeron trajectory~\cite{regge}. In 
the same framework, events in which the hadron loses a somewhat higher 
fraction of its energy are ascribed to the exchange of Reggeon and pion 
trajectories.

Significant progress has been made in understanding diffraction in terms 
of perturbative Quantum Chromodynamics (pQCD) by studying the dissociation 
of virtual photons, $\gamma^{\star}p \to Xp$, in diffractive deep 
inelastic $ep$ scattering (DIS) at HERA, $ep \to eXp$. The part of the DIS 
cross section due to such processes may be expressed in terms of the 
diffractive parton distribution functions (PDFs) of the proton. 
Diffractive PDFs are defined as the proton PDFs probed when the proton 
emerges intact from the hard interaction, suffering only a small energy 
loss.

At high centre-of-mass energy, diffractive $ep$ scattering is 
characterised by the presence of a leading proton in the final state 
carrying most of the proton beam energy and by the presence 
of a large rapidity gap (LRG) in the forward (proton) direction. Both of 
these signatures have been exploited at the HERA collider to select 
samples enriched in diffractive events. Alternatively, a method has been 
used to determine statistically the number of diffractive events, based on 
the expected difference in shape of the distributions of the invariant 
mass, $M_X$, for diffractive and non-diffractive events. These approaches 
are subject to different systematic uncertainties.

This paper presents results based on the detection of a leading proton or 
of a large rapidity gap. The same data have also been analysed in terms of 
the shape of the $M_X$ distribution~\cite{mx2}. For the proton-tagged 
sample, the ZEUS leading proton spectrometer (LPS) was used; this data 
sample has events with scattered protons carrying a fraction, $x_L$, of at 
least 90\% of the incoming proton momentum. For $x_L \lsim 0.98$--$0.99$, 
the 
sample is dominated by non-diffractive events, whilst for $x_L \approx 1$, 
it consists almost exclusively of diffractive events; therefore the 
transition between non-diffractive and diffractive regions is studied. In 
the LRG sample the proton momentum is not measured, but events are 
selected on the basis of the variable $\xpom$, which is the fraction of 
the proton's momentum carried by the diffractive exchange, $\xpom \simeq 
1-x_L$. Events in the LRG sample are required to have $\xpom <0.02$ and 
thus the sample mainly ($\gsim 90\%$) consists of diffractive 
events~\cite{GolecBiernat:1997vy}. The kinematic regions covered by the 
LRG and LPS results are: photon virtualities $~2~<~Q^2~<~305$\,GeV$^2$ 
(LRG) or $2<Q^2<120$\,GeV$^2$ (LPS), photon-proton centre-of-mass energies 
$40<W<240$\,GeV, hadronic final-state masses $2<M_X<25$\,GeV (LRG) or 
$2<M_X<40$\,GeV (LPS), proton fractional momentum losses 
$0.0002<\xpom<0.02$ (LRG) or $0.0002<\xpom<0.1$ (LPS) and values of the 
square of the four-momentum exchanged at the proton vertex 
$0.09<|t|<0.55$\,GeV$^2$ (LPS).

\section{Experimental set-up}

The data used for this measurement were taken with the ZEUS detector at 
the HERA $ep$ collider in the years 1999 and 2000, when HERA collided 
positrons of 27.5\,GeV with protons of 920\,GeV. The data used for the LRG 
and LPS analyses correspond to integrated luminosities of 62.2 pb$^{-1}$ 
and 32.6 pb$^{-1}$, respectively. 

A detailed description of the ZEUS detector can be found
elsewhere~\cite{bluebook,pl:b293:465}. A brief outline of the components
that are most relevant for this analysis is given below.

Deep inelastic scattering events were identified using information from 
the central tracking detector (CTD), the uranium--scintillator calorimeter 
(CAL), the small angle rear tracking detector (SRTD), the rear part of 
the hadron-electron separator (RHES) and the forward plug calorimeter 
(FPC).

Charged particles were tracked in the 
CTD~\cite{nim:a279:290,*npps:b32:181,*nim:a338:254}. The CTD operated in a 
magnetic field of 1.43 T provided by a thin solenoid. It consisted of 72 
cylindrical drift chamber layers, organised in nine superlayers covering 
the polar-angle\footnote{The ZEUS coordinate system is a right-handed 
Cartesian system, with the $Z$ axis pointing in the proton direction, 
referred to as the ``forward direction'', and the $X$ axis pointing left 
towards the centre of HERA. The coordinate origin is at the nominal 
interaction point.} region $15^\circ<\theta<164^\circ$.  The 
transverse-momentum resolution for full-length tracks was $\sigma(p_T)/p_T 
= 0.0058p_T \oplus 0.0065 \oplus 0.0014/p_T$, with $p_T$ in GeV.

The CAL~\citeCAL consisted of three parts: the forward (FCAL), the barrel
(BCAL) and the rear (RCAL) calorimeters. Each part was subdivided
transversely into towers and longitudinally into one electromagnetic
section (EMC) and either one (in RCAL) or two (in BCAL and FCAL) hadronic
sections (HAC). The smallest subdivision of the calorimeter was called a
cell.  The CAL energy resolutions, as measured under test-beam conditions,
were $\sigma(E)/E=0.18/\sqrt{E}$ for electrons and 
$\sigma(E)/E=0.35/\sqrt{E}$ 
for hadrons, with $E$ in~GeV. 

The position of electrons scattered at small angles to the
electron-beam direction was determined by means of the information from
the CAL and the SRTD~\cite{nim:a401:63,epj:c21:443}. The SRTD was
attached to the front face of the RCAL and consisted of two planes of
scintillator strips, 1\,cm wide and 0.5\,cm 
thick, arranged in orthogonal
orientations. Ambiguities in SRTD hits were resolved with the help of
the RHES~\cite{nim:a277:176}, which consisted of a layer of
approximately 10,000 ($2.96 \times 3.32$\,cm$^2$) silicon-pad detectors
inserted in the RCAL at a depth of 3.3 radiation lengths.

The FPC~\cite{nim:a450:235} was used to measure the energy of particles in 
the pseudorapidity range $\eta~\approx~4.0 - 5.0$. It was a 
lead--scintillator sandwich calorimeter read out by wavelength-shifter 
(WLS) fibers and photomultipliers (PMT). It was installed in the $20 
\times 20$\,cm$^2$ beam hole of the FCAL. The FPC had outer dimensions of 
19.2 $\times$ 19.2 $\times$ 108\,cm$^3$ and had a central hole of 3.15\,cm 
radius to accommodate the beam-pipe. In the FPC, 15~mm thick lead plates 
alternated with 2.6 mm thick scintillator layers.  The FPC was subdivided 
longitudinally into an electromagnetic (10 layers) and a hadronic section 
(50 layers) representing a total of 5.4 nuclear absorption lengths.  The 
energy resolution for electrons, as measured in a test beam, was 
$\sigma(E)/E = (0.41 \pm 0.02)/\sqrt{E} \oplus 0.062 \pm 0.002$, with $E$ 
in GeV. When installed in the FCAL, the energy resolution for pions was 
$\sigma(E)/E = (0.65 \pm 0.02)/\sqrt{E} \oplus 0.06 \pm 0.01$, with $E$ in 
GeV, and the $e/h$ ratio was close to unity.

The LPS~\cite{lps} detected positively charged particles scattered at 
very small angles and carrying a substantial fraction, $x_L$, of the 
incoming proton momentum; these particles remained in the beam-pipe and 
their trajectories were measured by a system of silicon microstrip detectors 
that could be inserted very close (typically a few mm) to the proton 
beam. The detectors were grouped in six stations, S1 to S6, placed along 
the beam line in the direction of the proton beam, between 23.8\,m and 
90.0\,m from the interaction point. The particle deflections induced by 
the magnets of the proton beam line allowed a momentum analysis of the 
scattered protons. Only stations S4, S5 and S6 covered the kinematic 
region of the present measurement. The resolutions were about $0.5\%$ 
on the longitudinal momentum 
fraction and about 5 MeV on the transverse momentum. The effective 
transverse-momentum resolution was dominated by the intrinsic 
transverse-momentum spread of the proton beam at the interaction point, 
which was about 45~MeV in the horizontal plane and about 100~MeV in the 
vertical plane. The LPS acceptance was approximately 2\% and 
$x_L$ independent for $x_L~\gsim~0.98$; it increased smoothly to about 
10\% as $x_L$ decreased to 0.9.

The luminosity was determined from the rate of the bremsstrahlung process 
$ep~\rightarrow~e\gamma p$. The photon was measured in a 
lead--scintillator calorimeter~\cite{lumi1,*lumi2,*lumi3} placed in the 
HERA tunnel at $Z=-107$~m.

\section{Kinematics and cross sections}
\label{sec-kin}
\begin{figure}[p]
\vfill
\begin{center}
\includegraphics[width=9cm,height=9cm]{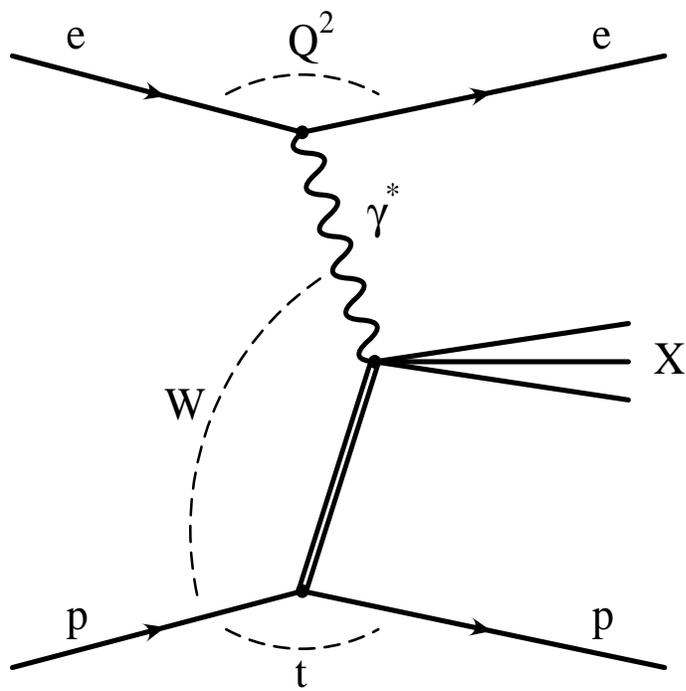}
\end{center}
\caption{
Schematic diagram of the reaction $ep \rightarrow eXp$.
}
\clearpage
\label{fig-contfey}
\vfill
\end{figure}

Figure~\ref{fig-contfey} shows a schematic diagram of the  
process
$ep \to e X p$. The kinematics of this reaction is described by
the following variables:
\begin{itemize}
\item
$Q^2=-q^2=-(k-k')^2$, the negative four-momentum squared of the virtual 
photon ($\gamma^{\star}$), where $k$ $(k')$ is the four-momentum of the 
incident (scattered) 
positron;
\item
$W^2=(q+P)^2$, the squared centre-of-mass energy of the photon-proton system, 
where $P$ is the four-momentum of the incident proton;
\item 
$x=Q^2/(2P\cdot q)$, the fraction of the proton 
momentum carried by the struck quark in the 
infinite-momentum frame (the Bjorken variable);
\item
$y=(q\cdot p)/(k \cdot p)$, the fraction of the positron energy 
transferred to the proton in the proton rest frame; 
\item
$M^2_X=(q+P-P')^2$, the squared mass of the system $X$,
where $P'$ is the four-momentum of the scattered proton;
\item
$t=(P-P')^2$, the squared four-momentum transfer at the proton vertex;
\item
$\Phi$, the angle between the positron scattering plane and the proton
scattering plane in the $\gamma^{\star}p$ centre-of-mass frame.
\end{itemize}

\noindent The variables $Q^2$, $W$ and $x$ are related by 
$x=Q^2/(Q^2+W^2-M_p^2)$, where $M_p$ is the proton mass.

The two
dimensionless variables $x_{\pom}$ and $\beta$ can be used instead of 
$M_X$ and $W$; they are given by
\begin{equation}
x_{\pom}=\frac{(P-P')\cdot q}{P\cdot q} = \frac{Q^2+M_X^2-t}{Q^2+W^2-M_p^2}~,
\label{eq-xpom}
\end{equation}
\begin{equation}
\beta=\frac{Q^2}{2(P-P')\cdot q} = \frac{Q^2}{Q^2+M_X^2-t}~.
\label{eq-beta}
\end{equation}
They are related to $x$ by 
$x_{\pom} \beta = x$. The variable $\beta$
is the Bjorken variable defined with respect to the four-momentum of the
exchanged object. The variable $x_{\pom}$ is often referred to as $\xi$ at 
hadron colliders.

The cross section for the reaction $ep \to eXp$ can be expressed in terms 
of the diffractive structure function $F_2^{D(4)}$ or of the reduced 
diffractive cross-section $\sigma_r^{D(4)}$, which are defined by the 
equation
\begin{eqnarray}
\frac{d\sigma^{ep \rightarrow eXp}}{
d\beta dQ^2dx_{\pom}dt} & = &
\frac{4\pi\alpha^2}{\beta Q^4}\biggl[1-y+\frac{y^2}{2(1+R^{D})}\biggr]
F_2^{D(4)}(\beta,Q^2,x_{\pom},t)\nonumber\\
& = & 
\frac{4\pi\alpha^2}{\beta Q^4}\biggl[1-y+\frac{y^2}{2}\biggr]
\sigma_r^{D(4)}(\beta,Q^2,x_{\pom},t)~.
\label{sigma-2}
\end{eqnarray}

\noindent The quantity $R^D= \sigma_L^{\gamma^{\star} p \rightarrow 
Xp}/\sigma_T^{\gamma^{\star} p \rightarrow Xp}$ is the ratio of the cross 
sections for longitudinally and transversely polarised virtual photons. 
The diffractive longitudinal structure function, $F_L^D$, is related to 
$R^D$ via $F_L^D=F_2^D R^D/(1+R^D)$. The diffractive reduced cross 
section and the diffractive structure function coincide if $R^D=0$. Since 
$R^D$ has not been measured, the results are presented in terms of the 
diffractive reduced cross section.

The structure-function $F_2^{D(3)}(\beta, Q^2,\xpom)$ and the reduced 
cross-section $\sigma_r^{D(3)}(\beta, Q^2,\xpom)$ are obtained by 
integrating $F_2^{D(4)}$ and $\sigma_r^{D(4)}$ over $t$,
\begin{eqnarray}
F_2^{D(3)}(\beta, Q^2,\xpom)& = & \int{F_2^{D(4)}(\beta,Q^2,x_{\pom},t) 
dt}~, \nonumber \\ \nonumber
\sigma_r^{D(3)}(\beta, 
Q^2,\xpom)& = &\int{\sigma_r^{D(4)}(\beta,Q^2,x_{\pom},t) dt}~.
\label{f2d3}
\end{eqnarray}
\noindent

The $\Phi$ dependence of the cross section is 
sensitive to the interference between the longitudinal and transverse 
amplitudes; the sensitivity to these interference terms disappears when 
$\Phi$ is integrated 
over. For unpolarised positrons and protons, the cross section can
be decomposed as
\begin{equation}
\frac{d\sigma^{ep\rightarrow eXp}}{d\Phi} \propto
\sigma_T^{\gamma^{\star}p \rightarrow Xp}+
\epsilon\sigma_L^{\gamma^{\star}p \rightarrow Xp}-
2\sqrt{\epsilon(1+\epsilon)}\sigma^{\gamma^{\star}p \rightarrow 
Xp}_{LT}\cos{\Phi}-
\epsilon\sigma^{\gamma^{\star}p \rightarrow Xp}_{TT}\cos{2\Phi}~,
\label{fullsigma}
\end{equation}
where $\sigma^{\gamma^{\star}p \rightarrow Xp}_{LT}$ is due to the 
interference term
between the amplitudes for longitudinal and transverse polarisations of the 
virtual photon and
$\sigma^{\gamma^{\star}p \rightarrow Xp}_{TT}$ is due to 
the interference term between
the amplitudes for the two transverse polarisations. The parameter 
$\epsilon$ is defined as $\epsilon = 2(1-y)/[1+(1-y)^2]$.

\section{Methods of selecting diffraction}
\label{sec:sel-diffr}

The kinematic properties of diffractive DIS, $ep\rightarrow eXp$, imply 
the following for the final-state proton and the hadronic system $X$: 

\begin{itemize}

\item the proton suffers only a small perturbation and emerges from the 
interaction carrying a large fraction, $x_L$, of the incoming proton 
momentum. Diffractive events appear as a peak at $x_L\approx 1$, the 
diffractive peak, which at HERA extends down to $x_L$ of about 
$0.98$~\cite{LPS97}. The absolute value of the four-momentum-transfer 
squared, $|t|$, is typically smaller than 1\,GeV$^2$, with $\langle 
|t| \rangle 
\approx 0.15$\,GeV$^{2}$~\cite{LPS97};

\item the difference in rapidity between the outgoing proton and the 
system $X$ is $\Delta \eta \approx \ln{(1/\xpom)}$~\cite{rev}. Since the 
cross section increases with decreasing $\xpom$, most of the events have 
small $\xpom$ and therefore a large separation in rapidity between the 
outgoing proton and any other hadronic activity in the event is expected;

\item conservation of momentum implies that the system $X$ must have a 
small mass ($M_X$) with respect to the photon-proton centre-of-mass 
energy, since $1-x_L~\gsim~M_X^2/W^2$.

\end{itemize}

Conversely, in non-diffractive DIS, both the hadronic system associated 
with the struck quark, which is largely measured in the detector, and that 
of the proton remnant, which peaks in the forward direction, originate 
from the hadronisation of colour-connected states. In this case, the 
distribution of the final-state particles is governed by conventional 
fragmentation and particles are emitted roughly uniformly in rapidity 
along the $\gamma^{\star} p$ axis. Rapidity gaps are thus expected to 
be exponentially suppressed~\cite{Derrick:1986xh}. 

Therefore, to select diffractive events, either the final-state proton 
can be detected (LPS method) or the different characteristics 
of the system $X$ in diffractive and non-diffractive events (hadronic 
methods) can be exploited.

In the hadronic methods, events with 
high $M_X$ are not accessible since the 
non-diffractive background grows with $M_X$ and the rapidity gap becomes 
more and more forward (and eventually becomes confined to the 
beam-pipe). In addition, 
the measured cross section includes a contribution from 
events of the type $ep~\rightarrow~eXN$, in which the proton also 
dissociates into a low-mass state, $N$, separated from $X$ by a rapidity 
gap. Hadrons from the system $N$ can either escape undetected into the 
forward beam-pipe or fall into the detector acceptance. 
The mass of the system $N$ thus enters as an additional variable, 
and the observed 
particles must be assigned either to $N$ or to $X$. The 
contribution from proton-dissociative events needs to be estimated from 
a Monte 
Carlo (MC) simulation.  While these limitations add to the systematic 
uncertainties, the statistical precision of the results is good due to 
the high acceptance of the central detector. Although the acceptance is 
not limited in $t$, no measurement of $t$ is possible because of the poor 
resolution of the transverse momentum of the system $X$.

High-$x_L$ samples selected by the LPS method have little or no background 
from proton-dissociative events or from non-diffractive DIS. They also 
allow a direct measurement of the variables $t$, $\Phi$ and $\xpom$, and 
give access to higher values of $M_X$.  The statistical precision, 
however, is poorer than for the results obtained with the hadronic methods 
because of the small acceptance of the proton taggers -- approximately 2\% 
in the diffractive-peak region.

At HERA, several diffractive analyses based on the scattered proton 
measurement have been carried out~\cite{lps94, lps95, LPS97, h1-fps}. 
Analyses with the hadronic methods have been performed with event 
selections based both on the presence of forward large rapidity gaps 
(LRG method)~\cite{h1-lrg,recent_review} and on the shape of the 
mass distribution of the system $X$ ($M_X$ method)~\cite{lps95, mx1, 
mx2}. 

The results presented here are based both on the LPS method and on the LRG 
method; the same data have also been analysed with the $M_X$ 
method~\cite{mx2}. No attempt is made to isolate the Pomeron contribution, 
i.e. diffraction in the strict sense of the term, from contributions due 
to Reggeon and pion trajectories or non-diffractive DIS: only the 
contribution of proton-dissociative events is statistically subtracted.

\section{Reconstruction of the kinematic variables}
\label{sec-rec}

The identification of the scattered positron was based on a neural 
network~\cite{sira} using information from the CAL. If the positron was at 
angles large enough to be inside the CTD acceptance, a CTD track was 
required.  The variables $W$ and $Q^2$ were reconstructed using a 
combination of the electron method and the double angle method~\cite{da}.
                                                                               
In the LPS analysis, the longitudinal ($p_Z$) and transverse ($p_X, 
p_Y$) momenta of the scattered proton were measured. The fractional 
energy of the outgoing proton, $x_L$, was defined as $x_L=p_Z/E_p$, where 
$E_p$ is the incoming proton energy. The variable $t$ is given by
\begin{equation}
t=-\frac{p_T^2}{x_L} -\frac{(1-x_L)^2}{x_L}M_p^2~,
\label{tpt2}
\end{equation}
where $p_T$ is the transverse momentum of the proton with respect to the
incoming beam direction. The $t$ resolution was approximately 
$\sigma(t)/t=0.14~{\rm GeV} \sqrt{|t|}$, with $|t|$ in GeV$^2$, 
and was dominated
by the angular spread of the HERA proton beam.  The proton and the 
positron momenta were used to determine $\Phi$, the azimuthal angle 
between the positron and proton scattering planes in the $\gamma^{\star}p$ 
frame. The resolution in $\Phi$ was approximately 0.2~rad.
                                                                         
The four-momentum of the system $X$ was determined from both calorimeter and 
tracking information. The energy deposits in the CAL and the track 
momenta measured in the CTD were combined into energy flow objects 
(EFOs)~\cite{zeusdiff,gennady} to obtain the best momentum resolution. 
The EFOs were corrected for energy losses due to the  material 
of the detector. The mass $M_X$ was evaluated as

\begin{equation}
M_{X,{\rm EFO}}^2=\biggl(\sum{E_i}\biggr)^2-\biggl(\sum{p_{X,i}}\biggr)^2-
\biggl(\sum{p_{Y,i}}\biggr)^2-\biggl(\sum{p_{Z,i}}\biggr)^2~,\nonumber
\end{equation} 

\noindent where ($E_i$, $p_{X,i}$, $p_{Y,i}$, $p_{Z,i}$) is the momentum 
four-vector of the $i^{\rm th}$ EFO and the sum runs over all EFOs 
not assigned to the scattered positron. 

In the LPS analysis, the mass $M_X$ was also determined from the
outgoing proton momentum as reconstructed in the LPS,  

\begin{equation}
M_{X, {\rm LPS}}^2 \approx [1-x_L(1+x)] W^2~. \nonumber 
\end{equation} 

\noindent 
The best resolution on $M_X$ was obtained with $M_{X, {\rm EFO}}$ when 
$M_X$ was small and with $M_{X, {\rm LPS}}$ when $M_X$ was large; $M_X$ was 
therefore reconstructed as
\begin{equation} 
M^2_{X} = w_{\rm EFO} M^2_{X,{\rm EFO}} + w_{\rm LPS} M^2_{X,{\rm LPS}}~,
\label{mx} 
\end{equation} 
where the weights $w_{\rm EFO}$ and $w_{\rm LPS}$ 
are inversely proportional to 
the corresponding resolutions, and $w_{\rm EFO}+w_{\rm LPS}=1$. The 
resulting resolution was 
$\sigma(M_X)/M_X=0.35/\sqrt{M_X}~+~0.08$, with $M_X$ in GeV.
                                                                               
The variables $\xpom$ and $\beta$ were obtained from Eqs.~(\ref{eq-xpom})
and~(\ref{eq-beta}), using the measured values of $Q^2$, $W$, $M_X$ and 
neglecting $t$, since $|t| \ll Q^2, M_X^2$.
                                                                               
The variable $y$ was reconstructed as 
$y_{\rm JB}=\sum{\frac{(E_i-p_{Z,i})}{2E_e}}$, 
where the sum runs over all
EFOs not assigned to the scattered positron and $E_e$ is the energy of
the incident positron (``Jacquet-Blondel method''~\cite{jb}).

\section{Event selection}
\label{sec-sel}

The data used for the analysis were selected at the trigger 
level~\cite{bluebook,Smith:1994nx,*Smitha} by requiring the 
presence 
of a scattered positron in the CAL. The trigger selection of the LRG data 
also required that the energy deposited in the FPC be smaller than 20\,GeV. 
For the LPS data, a scattered proton was required in the LPS.

Offline, the following cuts were imposed:

\begin{itemize}

\item the energy of the scattered positron was required to be larger than 
10\,GeV. The position of the scattered positron was required to be within 
the fiducial region of the CAL. This was defined by a set of 
cuts~\cite{jarek-thesis} which removed regions where the inactive material 
was not adequately simulated or where the positron shower was not fully 
contained;

\item the requirement $45<(E-P_Z)<65$\,GeV was imposed. Here $E-P_Z = 
\sum{(E_i-p_{Z,i})}$, with the summation running over all EFOs including 
the scattered positron. This cut reduced the size of the QED radiative 
corrections and the photoproduction background, where the scattered 
positron escaped undetected in the rear beam hole;

\item the $Z$ coordinate of the interaction vertex, $Z_{\rm vtx}$, was 
required to be in the range $-50<Z_{\rm vtx}<50$\,cm. Events without a 
measured vertex were assigned to the nominal interaction point;

\item events with two electron candidates, of which at least one lacked an 
associated track, and which were back-to-back within 5$^\circ$ 
in the azimuthal plane, were rejected. This cut removed the contribution 
of QED Compton scattering and deeply virtual Compton scattering.

\end{itemize}

For the LRG sample, the presence of a rapidity gap of at least two 
units between the hadronic 
final-state $X$ and the outgoing proton was ensured by requiring that 
the energy deposited in the FPC, $E_{\rm FPC}$, be smaller than 1\,GeV
and by 
demanding $\eta_{\rm max}^{\rm CAL}<3$. Here $\eta_{\rm max}^{\rm CAL}$ 
is the pseudorapidity 
of the most forward EFO with energy above 400 MeV in the CAL. This 
combination of cuts suppressed background from non-diffractive and 
proton-dissociative processes.

The following requirements were used to select the scattered proton 
measured in the LPS:

\begin{itemize}

\item only events with $p_X<0$ were used as, for the present sample, the 
LPS acceptance for $p_X>0$ was low;

\item the candidate proton was tracked along the beam line and was 
rejected if the distance of closest approach to the beam-pipe was 
less than 0.2\,cm. It was also rejected if the $X$ position of the track 
impact point at station S4 (upper part) was smaller than $-3.0$\,cm. These 
cuts reduced the sensitivity of the acceptance to the uncertainty in the 
position of the beam-pipe apertures;
                                                                          
\item

beam-halo background was caused by scattered protons with energy close to
that of the beam, originating from the interaction of a beam proton with
the residual gas in the beam-pipe or with the collimators. A
beam-halo proton may overlap with a standard non-diffractive DIS event. In
this case, the proton measured in the LPS was uncorrelated with the
activity in the central detector. This background was suppressed by
the requirement that the sum of the energy and the longitudinal component
of the total momentum measured in the CAL and the LPS be less
than the kinematic limit of twice the incoming proton energy:
$E+P_Z \simeq (E+P_Z)_{\rm CAL} +2 p_Z^{\rm LPS} <1860$\,GeV.
This cut took into account the resolution of the measurement of 
$p_Z^{\rm LPS}$. 
The residual beam-halo background and its subtraction are discussed
in Section~\ref{sec-lpsbkg}.

\item
the variable $t$ was required to be in the range $0.09<|t|<0.55$\,GeV$^2$.
This cut eliminated regions where the LPS acceptance was small or rapidly
changing.

\end{itemize}
                                                                               
The LRG analysis was further restricted to the regions 
$2<Q^2<305$\,GeV$^2$, $40<W<240$\,GeV, $2<M_X<25$\,GeV and 
$0.0002<x_{\pom}<0.02$; the average $Q^2$ value is 13\,GeV$^2$. For the 
LPS sample, the region selected was 
$2<Q^2<120$\,GeV$^2$, $40<W<240$\,GeV, $2<M_X<40$\,GeV and 
$0.0002<x_{\pom}<0.1$; the average $Q^2$ value for the LPS sample is 
11\,GeV$^2$. These selections yielded 708,851 events for the LRG analysis 
and 15,130 for the LPS analysis.

The LRG and LPS samples were collected 
simultaneously: 0.7\% of the LRG events have a proton measured in the LPS 
and 35\% of the LPS events are also contained in the LRG sample.

\section{Monte Carlo simulation and acceptance corrections}
\label{sec-mc}

Monte Carlo simulations were used to correct the data for
acceptance and detector effects.

Diffractive events were simulated with the {\sc Satrap} 
generator \cite{satrap}, 
which is based on the saturation model of Golec-Biernat and 
W\"usthoff~\cite{gbw1,*gbw2,*gbw3}. 
{\sc Satrap} is embedded in the {\sc Rapgap} framework~\cite{rapgap}. 
The version of 
{\sc Satrap} used here is identical to that used in a previous ZEUS 
publication~\cite{mx2}, except for some 
reweighting to describe the measured distributions for the higher-$\xpom$ 
events, where Reggeon and pion exchanges become significant.

Diffractive events were also modelled with {\sc Rapgap} 2.08/06~\cite{rapgap}, 
which is based on the model of Ingelman and Schlein~\cite{ingelman} and 
assumes Regge factorisation: 
the structure function $F_2^{D(4)}$ is expressed as the sum of 
separately factorisable Pomeron and Reggeon contributions, 
\begin{equation}
F_2^{D(4)}(x_{\pom},t,\beta,Q^2)=
f_{\pom}(x_{\pom},t)F_2^{\pom}(\beta,Q^2)+f_{\reg}(x_{\pom},t)F_2^{\reg}(\beta,Q^2)~.
\label{i-s}
\end{equation}
The Pomeron and Reggeon fluxes, $f_{\pom, \reg}(x_{\pom},t)$,
were parameterised~\cite{regge} as
\begin{equation}
f_{\pom, \reg}(x_{\pom},t)=\frac{e^{b_0^{\pom,
\reg}t}}{x_{\pom}^{2\alpha_{\pom,\reg}(t)-1}}~,
\label{flux}
\end{equation}
with linear trajectories $\alpha_{\pom, \reg}(t)=\alpha_{\pom, 
\reg}(0)+\alpha_{\pom, \reg}'t$, and with the values of the parameters 
taken from hadron-hadron data~\cite{softpom}. The Pomeron 
structure function 
$F_2^{\pom}(\beta,Q^2)$ was taken from the H1 
dPDF fit 2~\cite{h1diff}. 
The structure function $F_2^{\reg}$ is unknown and was 
assumed to be that of the pion~\cite{pion}. 

The process of QCD radiation was simulated differently in the two MC 
samples. In the {\sc Satrap} sample, a parton-shower model as implemented 
in {\sc Meps}~\cite{Bengtsson:1987rw} was used. In the {\sc Rapgap} 
sample, higher-order QCD radiation was simulated with {\sc 
Ariadne}~\cite{Lonnblad:1992tz}. In both cases hadronisation was simulated 
with the Lund string model as implemented in {\sc 
Jetset}~7.4~\cite{Bengtsson:1987kr,Sjostrand:1993yb}.

Initial- and final-state QED radiation was simulated by using {\sc Satrap} or 
{\sc Rapgap} in conjunction with {\sc Heracles} 
4.6~\cite{heracles}. The measurements were 
corrected for these effects and the cross sections are presented at the 
Born level.

The inclusive DIS events were simulated with {\sc Djangoh} 1.1~\cite{django}, 
using the CTEQ4D \cite{cteq4} parameterisation of the proton parton 
densities.

The {\sc Pythia} 6.2 generator~\cite{pythia} 
was used to study the photoproduction background as well as the 
proton-dissociative contribution (see Section~\ref{sec-bkglrg}). 
Events in the proton-dissociative 
MC sample were reweighted such as to give a good description of all 
measured variables in the data. 

\begin{figure}[p]
\vfill
\begin{center}
\includegraphics[width=15cm,height=15cm]{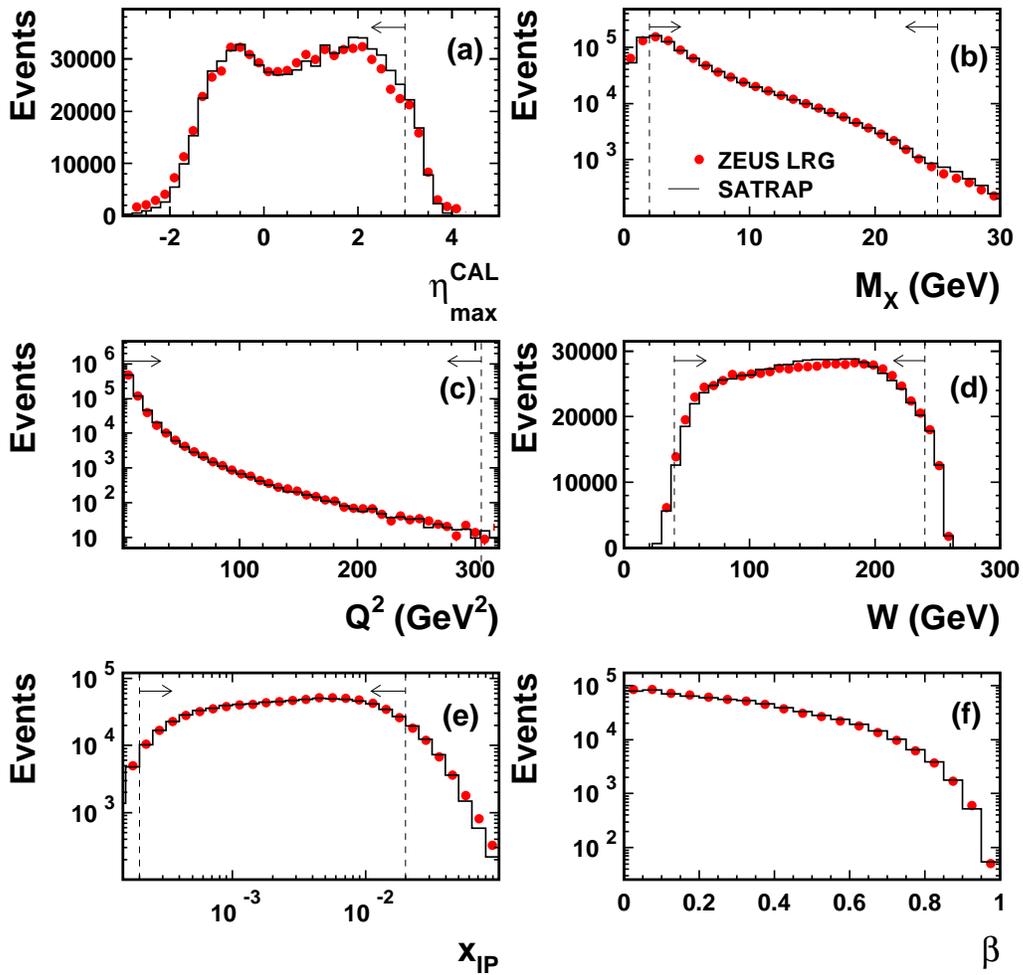}
\end{center}
\caption{
Comparison of the distributions measured (dots) and obtained 
with the reweighted {\sc Satrap} MC (histograms)
for (a) $\eta_{\rm max}^{\rm CAL}$, (b) $M_X$, (c) $Q^2$, (d) $W$,  
(e) $x_{\pom}$ and (f) $\beta$ in the LRG analysis. 
For each plot, all analysis cuts have been applied except that on the
plotted variable. The vertical lines with arrows indicate the selected 
region. 
} 
\label{fig-data-mc-lrg}
\vfill
\end{figure}

\begin{figure}[p]
\vfill
\begin{center}
\includegraphics[width=15cm,height=15cm]{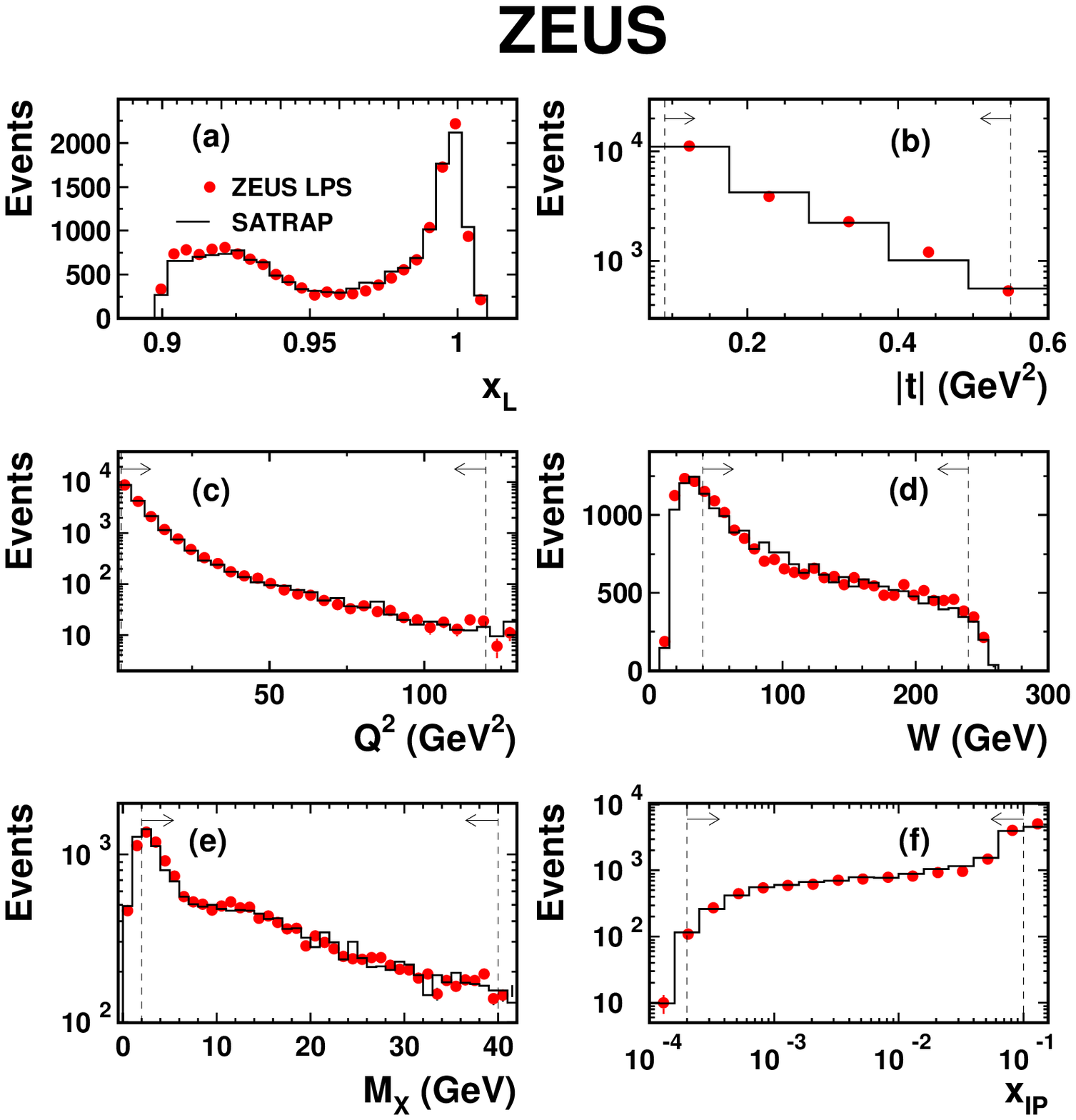}
\end{center}
\caption{
Comparison of the distributions measured (dots) and obtained 
with the reweighted {\sc Satrap} MC (histograms)
for (a) $x_L$, (b) $|t|$, (c) $Q^2$, (d) $W$, (e) $M_X$ and (f) 
$x_{\pom}$ in the LPS analysis. 
Other details as in caption for Fig.~\ref{fig-data-mc-lrg}.
}
\label{fig-data-mc-lps}
\vfill
\end{figure}

All generated events were passed through the standard ZEUS detector 
simulation, based on the {\sc Geant} 3.13 
program~\cite{geant}, and through the 
trigger simulation package. The measurements were corrected for detector 
acceptance and resolution, and for radiative effects, with suitable 
combinations of the various MC models. 
A comparison of data and SATRAP for the LRG 
analysis is presented in Fig.~\ref{fig-data-mc-lrg} for the variables 
$\eta_{\rm max}^{\rm CAL}$, $M_X$, $Q^2$, $W$, $x_{\pom}$ and $\beta$. The 
simulation is in satisfactory agreement with the data in the region of 
interest, indicated by vertical lines in the plots. 
A similar comparison 
for the LPS analysis is presented in Fig.~\ref{fig-data-mc-lps} for the 
variables $x_L$, $|t|$, $Q^2$, $W$, $M_X$, and $x_{\pom}$. The simulation 
reproduces the data reasonably well. The diffractive peak is evident in 
Fig.~\ref{fig-data-mc-lps}a.

\subsection{Cross-section extraction}
\label{sec-xsec}

The diffractive reduced cross section at a given point within a bin was 
obtained from the ratio of the background-subtracted number of events to 
the number of events in that bin predicted by SATRAP, 
multiplied by the Born-level reduced cross section used in SATRAP. Both 
the acceptance and the bin-centring corrections were 
thus taken from SATRAP.

For the LPS data, the cross section was directly measured only in a 
limited $t$ region and extrapolated to $0<|t|<1$\,GeV$^2$ assuming an 
exponential $t$-dependence, $d\sigma^{ep \rightarrow eXp}/dt \propto 
\exp{(-b|t|)}$, with $b= 7.0$\,GeV$^{-2}$. The effect of the 
extrapolation is to increase the cross section by a factor of about two; 
this factor is largely independent of kinematics. 
Data from elastic and proton-dissociative $pp$ and $\bar{p}p$ scattering 
indicate that the $t$ distribution is better described by the function 
$\exp{(-b|t|+ct^2)}$. For example, fits to the $\bar{p}p$ data at 
$\sqrt{s}=546$\,GeV~\cite{sps} yield $c=2.3 \pm 0.1$\,GeV$^{-4}$. In the 
extrapolation to the range $0<|t|<1$\,GeV$^2$, $c$ was nominally set to 
zero and changed up to 4\,GeV$^{-4}$, yielding changes in the 
extrapolated cross section of up to $+9\%$ (setting $c$ to 2~GeV$^{-4}$ 
changes the cross section by $+6\%$). This effect was included in the 
normalisation uncertainty discussed in Section~\ref{sec-sys}.

\section{Backgrounds}
\label{sec-bkg}

\subsection{LRG analysis}
\label{sec-bkglrg}

The main source of background in the LRG sample comes from events of the 
type $ep~\rightarrow~eXN$, in which the proton dissociates into a 
low-mass system, $N$. The proton-dissociative system can either escape 
entirely undetected in the forward beam-pipe or leak partially into the 
detector acceptance and therefore be measured by the FPC and the CAL. In 
the former case, the background events are included in the measured cross 
section, so that they bias the normalisation. As indicated by MC 
simulations, in the latter case most of the events are rejected by the 
FPC veto and by the $\eta_{\rm max}^{\rm CAL}$ cut.

The ratio of the LPS to the LRG results measures the fraction of 
proton-dissociative events in the LRG sample.  The ratio is $0.76 \pm 
0.01 {\rm (stat.)} ^{+0.03} _{-0.02} {\rm (syst.)} ^{+0.08} _{-0.05} {\rm 
(norm.)} $ and is independent of $Q^2$, $\xpom$ and $\beta$, as discussed 
in Section~\ref{sec-f2d3}; the last uncertainty reflects the 
normalisation uncertainties, mostly due to the LPS data. The percentage 
of proton-dissociative events in the LRG sample is therefore $24 \pm 1 
{\rm (stat.)} ^{+2} _{-3} {\rm (syst.)} ^{+5} _{-8}{\rm (norm.)} $\%.

The contribution of proton-dissociative events to the measured 
cross sections was also estimated with {\sc Pythia}. A sample of 
proton-dissociative data was selected in two alternative ways:

\begin{itemize} 

\item by requiring $\eta_{\rm max}^{\rm CAL}<2$ and $E_{\rm FPC}>1$\,GeV, 
and the remaining selection as described in 
Section~\ref{sec-sel} for the LRG events 
(this will be referred to as FPC PDISS sample);

\item by adding to the selection described in Section~\ref{sec-sel} for 
the LRG events the requirement that a proton be measured in the LPS with 
$0.5<x_L<0.9$ (LPS PDISS sample).

\end{itemize}

\begin{figure}[p] \vfill \begin{center}
\includegraphics[width=15cm,height=15cm]{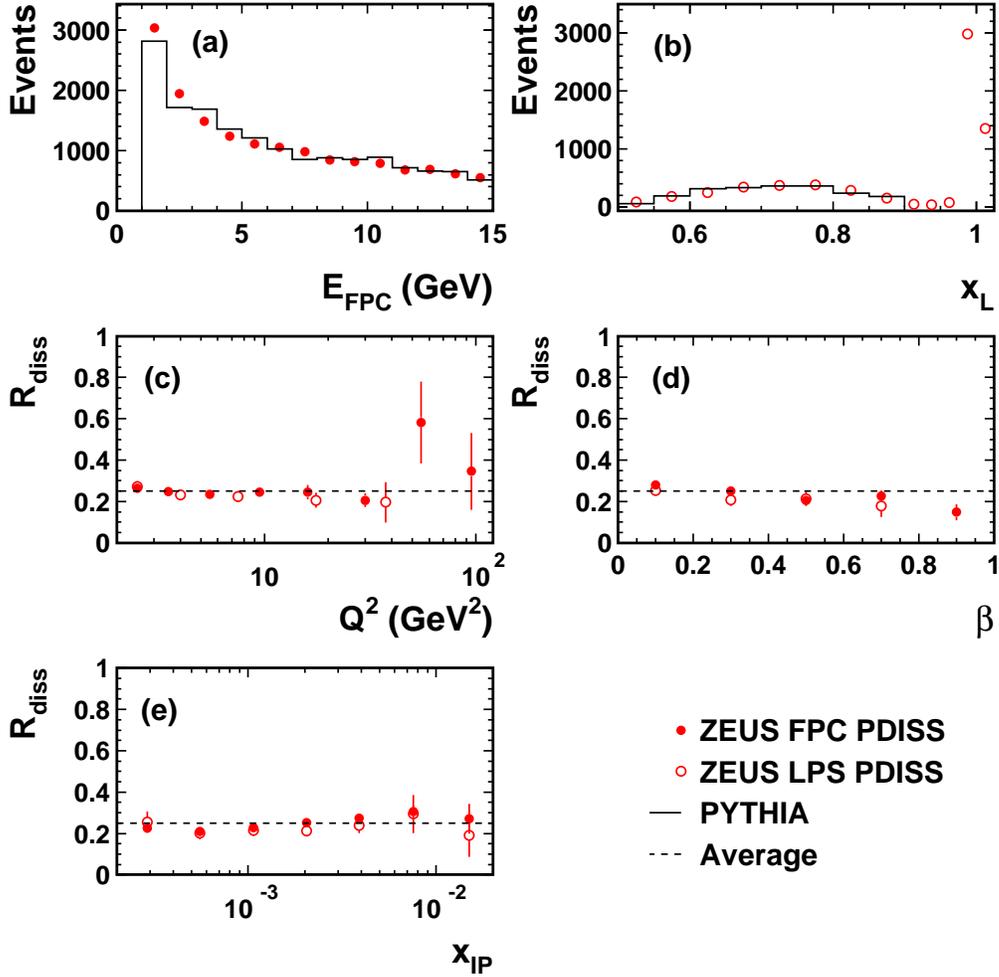} 
\end{center}
\caption{(a) The energy distribution in the FPC, $E_{\rm FPC}$,  
for the sample of proton-dissociative  
candidate events (FPC PDISS). The data (dots) are compared to the 
expectation of the reweigthed {\sc Pythia} MC (histogram), normalised to the 
data. (b) The $x_L$ distribution in the LPS for the 
sample of proton-dissociative 
candidate events (LPS PDISS). The data (open circles) are compared to the 
expectation of the reweigthed {\sc Pythia} MC (histogram), normalised to 
the data in the range $0.5 < x_L < 0.9$. The data points for $x_L > 0.9$ 
are also shown for completeness. The extracted fraction of 
proton-dissociative events, $R_{\rm diss}$, from the two samples as a 
function 
of (c) $Q^2$, (d) $\beta$ and (e) $\xpom$. The dashed lines in (c), (d) and 
(e) represent the average of the points.
} 
\label{fig-pdiss-fraction} 
\vfill
\end{figure}

The generated {\sc Pythia} distributions for $M_N$, $M_X$ and $Q^2$ were 
reweighted to give the best description of these data samples, in 
particular the $E_{\rm FPC}$ distribution in the FPC PDISS sample and 
the $x_L$ distribution in the LPS PDISS sample. The median of the 
generated $M_N$ distribution in {\sc Pythia} is 1.7\,GeV. The median of 
the same distribution for the events which pass the LRG analysis cuts is 
1.6\,GeV. Figures~\ref{fig-pdiss-fraction}a--b show the 
comparison of {\sc Pythia} with the proton-dissociative samples FPC and 
LPS PDISS as a function 
$E_{\rm FPC}$ and $x_L$, respectively. Also shown in 
Figs.~\ref{fig-pdiss-fraction}c--e is the fraction of proton-dissociative 
events expected in the LRG sample as a function of $Q^2$, $\beta$ and 
$x_{\pom}$. This fraction, obtained separately from the LPS and FPC PDISS 
samples, is constant at the level of $25$\% in both cases. 
The average of the FPC and LPS estimates provides a measurement of the 
proton-dissociative contribution to the LRG sample of $25 \pm 1 {\rm 
(stat.)} \pm 3 {\rm (syst.)}\%$, consistent with the ratio of the LPS to 
LRG results quoted above. The systematic uncertainty was estimated by 
varying the shape of the generated $M_N$ distribution, by changing the 
FPC cut as well as the $\eta_{\rm max}^{\rm CAL}$ cut and by taking into 
account the LPS normalisation uncertainty.  The combination of the LPS 
and FPC PDISS samples covers nearly the whole $M_N$ spectrum, including 
the lowest $M_N$ values. This fact, along with the agreement with the 
LPS to LRG 
ratio, lends support to the present estimate of the proton dissociation 
background. A background contribution of $R_{\rm diss}=25 \pm 1 {\rm 
(stat.)} \pm 3 {\rm (syst.)}\%$ was therefore subtracted from the 
data\footnote{In terms of the ratio $R_{M_X}=1/(1-R_{\rm diss})$ used 
elsewhere~\protect\cite{lps95,LPS97}, this background contribution 
corresponds to $R_{M_X}=1.33 \pm 0.02 {\rm (stat.)} \pm 0.05 {\rm (syst.)} 
$.}. Unless stated otherwise, all results are thus given for the reaction 
$ep \to eXp$, i.e. $M_N=M_p$.

The {\sc Pythia} generator was also used to evaluate the photoproduction 
background, which arises from low-$Q^2$ events in which the 
scattered positron escapes undetected in the rear direction and one of 
the final-state hadrons is misidentified as a positron. The largest 
contribution was found in the lowest $Q^2$ bin ($2<Q^2<3$\,GeV$^2$), where 
it was about 1.2\%. This background was neglected.

The contribution of non-diffractive events, estimated with {\sc Djangoh} 
1.1, was found to be roughly $10\%$ in the highest $\xpom$ bin ($0.01 < 
\xpom < 0.02$) and to decrease rapidly with decreasing $\xpom$. This 
background was not subtracted but bins in which the contribution was 
larger than $10\%$ were rejected.

\subsection{LPS analysis}
\label{sec-lpsbkg}

The main background contribution in the LPS sample at high $x_L$ is given 
by proton beam-halo events. In such events, the proton detected in the 
LPS is not correlated with the measurements in the central detector. To 
estimate this background, the variable $E+P_Z$ (see 
Section~\ref{sec-sel}) was used. For a signal event, this quantity should 
be equal to twice the initial proton energy, 1840\,GeV, whereas for a 
beam-halo event it can exceed this value.

\begin{figure}[p]
\vfill
\begin{center}
\includegraphics[width=15cm,height=15cm]{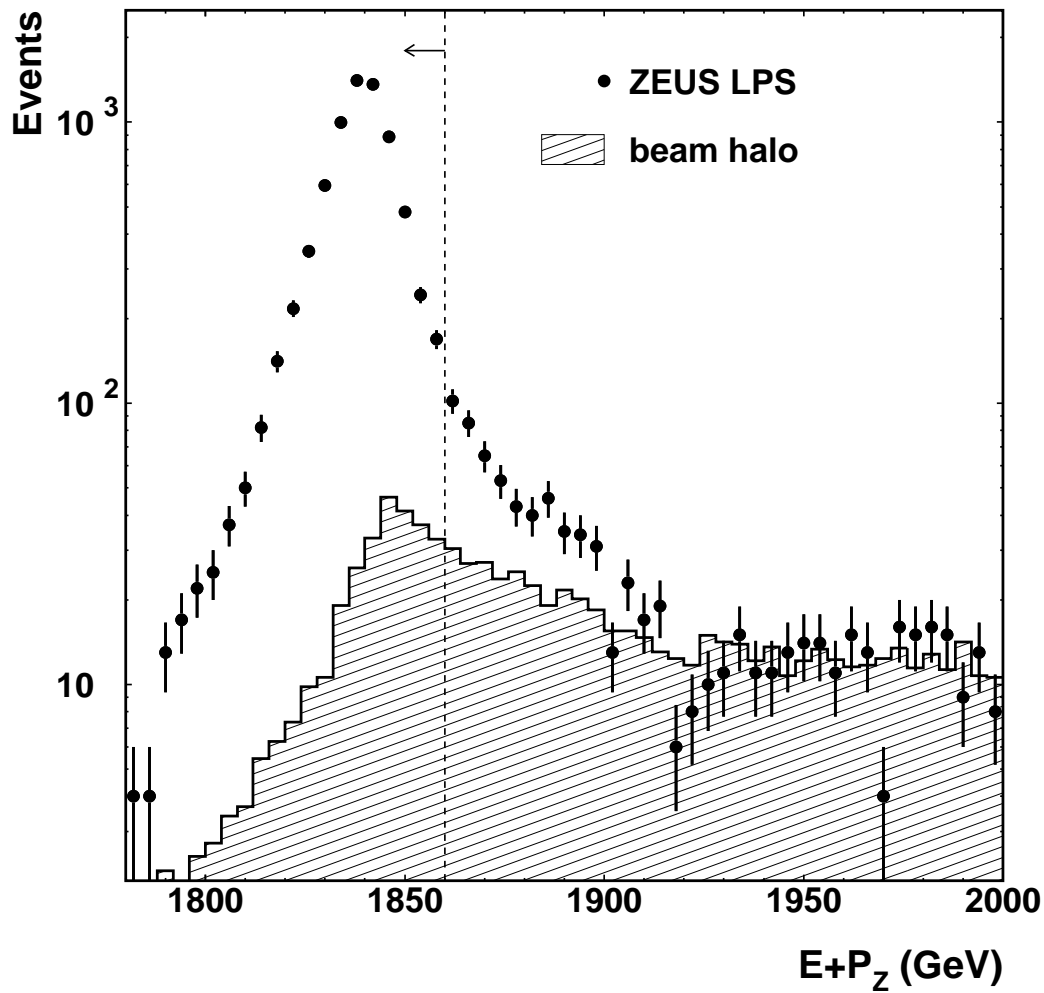} 
\end{center}
\caption{Distribution of $E+P_Z$ for the LPS events (dots). The dashed 
histogram represents the estimate of the beam-halo background normalised 
for $E+P_Z > 1925$~{\rm GeV}. The vertical dashed line at $E+P_Z = 
1860$~{\rm GeV} represents the selection cut used in the analysis. 
}
\label{fig-halo}
\vfill
\end{figure}

The $E+P_Z$ spectrum for the beam-halo events was constructed as a random 
combination of a generic DIS event (without the requirement of a track in 
the LPS) and a beam-halo track measured in the LPS, uncorrelated with the 
measurement in the main detector; here $P_Z$ includes the contribution of 
the energy deposition in the CAL and the proton momentum measured in the 
LPS. The resulting distribution, shown in Fig.~\ref{fig-halo} as the 
histogram, was normalised to the data for $E+P_Z>1925$\,GeV; this part of 
the distribution contains beam-halo events only. The background remaining 
after the cut at $E+P_Z<1860$\,GeV averages to $3.0 \pm 0.1~(\rm 
stat.)\%$, and is a decreasing function of $\xpom$. The 
results presented in this paper were corrected for this background.

The contribution from proton-dissociative events, $ep \to eXN$, studied 
with {\sc Pythia}, was around 9\% at $\xpom=0.1$, decreasing rapidly with 
decreasing $\xpom$. All results were corrected for this background.
In the region $\xpom <0.02$,  this background is negligible. 

The photoproduction background was negligible.

\section{Systematic uncertainties}
\label{sec-sys}

The systematic uncertainties were estimated~\cite{LPS97, jarek-thesis} 
by varying the cuts and by modifying the analysis 
procedure.  The variations of the cuts were typically commensurate with 
the resolutions of the relevant variables.

For each systematic check, the average effect on the cross section in the 
measured bins is indicated using the notation $(^{+a}_{-b})$. Given a 
systematic check which produced an increase of the cross section in some 
bins and a decrease in some other bins, $a$ is the average increase and 
$b$ is the average decrease. 

For both the LPS and LRG analyses, the following checks were performed:

\begin{itemize}

\item to evaluate the uncertainties due to the measurement of the 
scattered positron, the fiducial region for the impact position of the 
positron on the face of the CAL around the rear beam-pipe was enlarged by 
1\,cm ($^{+1.2}_{-0.2}$)\%;

\item the minimum energy of the positron was increased to 12\,GeV 
($^{+0.2}_{-0.3}$)\%; 

\item the minimum value of $E-P_Z$ was raised to 47\,GeV 
($^{+1.0}_{-0.7}$)\%;

\item the cut on the $Z$ coordinate of the vertex was restricted to 
$-40<Z_{\rm vtx}<40$\,cm~($^{+0.5}_{-0.5}$)\%;

\item the effect of the uncertainty in the absolute 
calorimeter energy calibration was estimated by changing the energy scale 
by $\pm2\%$ in the data only, separately for the scattered positron 
($^{+2.5}_{-2.3}$)\% and the hadronic system ($^{+2.3}_{-2.3}$)\%; 

\item the $x_{\pom}$ distribution in the MC was reweighted by a factor 
$(x_{\pom}/0.01)^k$, with $k$ varying between $-0.03$ and $+0.03$; the 
effect was ($^{+0.4}_{-0.4}$)\% in the LPS analysis and ($^{+3.0}_{-2.9}$)\% 
in the LRG analysis, the difference being mainly due to the correlation 
between the $\xpom$ and $\eta_{\rm max}$ variables.

\end{itemize}

For the LRG analysis, the following specific checks were also performed:

\begin{itemize}

\item the FPC energy cut was lowered to 0.7\,GeV ($^{+0.5}_{-0.4}$)\%;

\item the energy threshold on the most forward EFO used to reconstruct 
$\eta_{\rm max}^{\rm CAL}$ was lowered to 300 MeV ($^{+0.4}_{-0.3}$)\%
and increased to 500 MeV ($^{+0.2}_{-0.4}$)\%.

\end{itemize}

For the LPS analysis, the following specific checks were also performed:

\begin{itemize}

\item the cut on the minimum distance of approach to the beam-pipe was 
increased by 0.03\,cm ($^{+0.4}_{-0.8}$)\%;

\item the $t$ range was restricted to $0.1<|t|<0.5$\,GeV$^2$ 
($^{+4.1}_{-5.1}$)\%;

\item the proton-dissociative background was varied by $\pm30\%$ 
($^{+0.9}_{-0.9}$)\%;

\item the value of the $t$-slope in the MC was changed by 
$\pm1$\,GeV$^{-2}$ ($^{+4.0}_{-2.9}$)\%;

\item the $\Phi$ distribution in the MC was reweighted by a factor 
$(1+k\cos{\Phi})$, with $k$ varying between $-0.15$ and +0.15 
($^{+1.0}_{-0.9}$)\%;

\item the intrinsic transverse-momentum spread of the proton beam at the 
interaction point was increased by 5 MeV in the horizontal plane and 10 
MeV in the vertical plane ($^{+1.6}_{-1.9}$)\%.

\end{itemize}

The total systematic uncertainty for each bin was taken as the quadratic 
sum of the individual contributions. The effect of using the 
generator {\sc Rapgap} for the acceptance corrections instead of 
{\sc Satrap} was estimated ($^{+9.3}_{-8.6}$)\% but not included in the 
error bars as {\sc Rapgap} was found to provide a poor description of the data 
distributions.

For the LPS data, there is also an overall uncertainty of 
$\pm 7\%$ which originates mostly from the uncertainty of the simulation 
of the proton-beam optics -- largely independent of the kinematic 
variables, and therefore taken as a normalisation uncertainty. It also 
includes the uncertainty on the integrated luminosity ($\pm 2.25\%$).

In the LPS results integrated over $t$ ($\sigma_r^{D(3)}$ and 
$d\sigma/d\Phi$), an additional $+9\%$ uncertainty is present, due to the 
extrapolation from the measured to the full $t$ range (see Section 
\ref{sec-xsec}). The overall LPS normalisation uncertainty then becomes 
$^{+11}_{-7}\%$.

For the LRG data, the uncertainty on the integrated luminosity ($\pm 
2.25\%$) and that on the proton dissociation background ($\pm 4\%$) 
give an overall normalisation uncertainty of $\pm 5\%$.

\section{Results}
\label{sec-res}

The results in this section are presented as follows. The  
cross-sec\-tion $d\sigma^{ep\rightarrow eXp}/dt$ in the region 
$0.09<|t|<0.55$\,GeV$^2$ is discussed first. The data are then integrated 
over $t$ and extrapolated to the range $0<|t|<1$\,GeV$^2$, as discussed in 
Section~\ref{sec-xsec}. The resulting cross sections are presented as a 
function of $\Phi$ in Section~\ref{sec-phi}, where the sensitivity of the 
present data to the helicity structure of the reaction 
$ep~\rightarrow~eXp$ is discussed. The LPS data were used for both the 
$t$ and the $\Phi$ 
cross sections. In Sections~\ref{sec-f2d4} and~\ref{sec-f2d3}, the data 
are presented in terms of the diffractive reduced cross sections, 
$\sigma_r^{D(4)}$ and $\sigma_r^{D(3)}$. The former was measured, for the 
first time, in two bins of $t$, and was obtained from the LPS data. The 
latter was obtained both from the LPS data, after integration over $t$, 
and from the LRG data.
In Section~\ref{intercept}, the $\xpom$ dependence of $\sigma_r^{D(4)}$ 
and $\sigma_r^{D(3)}$ is used to extract the intercept of the Pomeron 
trajectory, $\alpha_{\pom}(0)$, the quantity that, in Regge phenomenology, 
determines the energy dependence of the total hadron-hadron cross 
section~\cite{regge}. 

The results for the LPS sample extend up to $\xpom = 0.1$. In this 
paper, the LPS data in the diffractive-peak region are often compared 
with those at high $\xpom$. For this purpose, the value $\xpom=0.01$ was 
chosen as the transition between the high- and low-$\xpom$ bins, such 
that the low-$\xpom$ bins are dominated by diffractive-peak events, while 
at higher $\xpom$ Reggeon and pion exchanges are 
important~\cite{GolecBiernat:1997vy}. This choice is somewhat 
restrictive, since the diffractive peak extends well below $x_L = 0.99$, 
see Fig.~\ref{fig-data-mc-lps}a. In the region $\xpom <0.01$, the 
contribution from non-Pomeron exchanges is less than 10\%. The average 
value of $\xpom$ is 0.003 for $\xpom<0.01$ and 0.043 for 
$0.01<\xpom<0.1$.

\subsection{\boldmath{$t$} dependence}
\label{sec-t}

\begin{figure}[p] \vfill \begin{center}
\includegraphics[width=15cm,height=15cm]{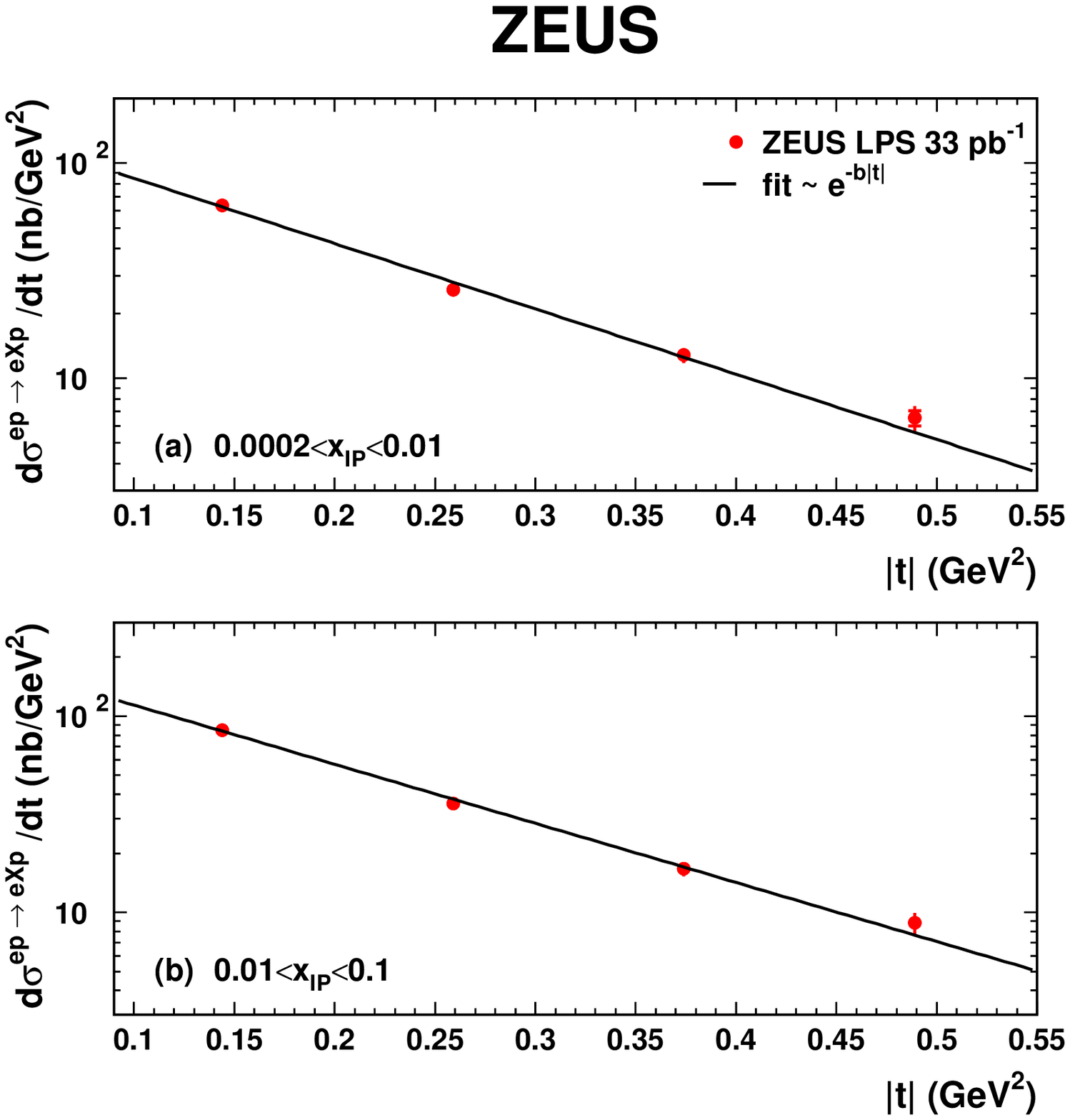} 
\end{center}
\caption{ 
The differential cross-section $d\sigma^{ep\rightarrow eXp}/dt$ 
for (a) $0.0002<x_{\pom}<0.01$ and (b) $0.01<x_{\pom}<0.1$. 
The lines show the results of fits with the function 
$d\sigma^{ep\rightarrow eXp}/dt \propto e^{-b|t|}$.
The inner error 
bars show the statistical uncertainties and the full bars indicate the 
statistical and the systematic uncertainties added in quadrature. The 
normalisation uncertainty of $\pm 7$\% is not shown. 
} 
\label{fig-tdistribution} \vfill
\end{figure}

The differential cross-section $d\sigma^{ep\rightarrow eXp}/dt$, obtained 
from the LPS data in the kinematic range $2<Q^2<120$\,GeV$^2$, 
$2<M_X<40$\,GeV, $40<W<240$\,GeV and $0.09<|t|<0.55$\,GeV$^2$, both for 
$0.0002<x_{\pom}<0.01$ (diffractive-peak region) and $0.01<x_{\pom}<0.1$, 
is presented in Fig.~\ref{fig-tdistribution} and 
Table~\ref{tab-tdistribution}.

The data were fitted with the single-exponential function 
$d\sigma^{ep\rightarrow eXp}/dt \propto e^{-b|t|}$. The value of the slope 
parameter, $b$, obtained from the fit in the region $0.0002<x_{\pom}<0.01$ 
is $b=7.0\pm 0.3~\rm{GeV^{-2}}$, with $\chi^2$/ndf=1.8 (ndf=2) when 
statistical and systematic uncertainties summed in quadrature are used in 
the fit. This result agrees with the previous ZEUS result~\cite{LPS97}. In 
the high-$\xpom$ region, $0.01<x_{\pom}<0.1$, the fit gives $b=6.9\pm 
0.3~\rm{GeV^{-2}}$ with $\chi^2$/ndf=1.1, again when the quadratic sum of 
statistical and systematic uncertainties is used.

\begin{figure}[p] \vfill \begin{center}
\includegraphics[width=15cm,height=15cm]{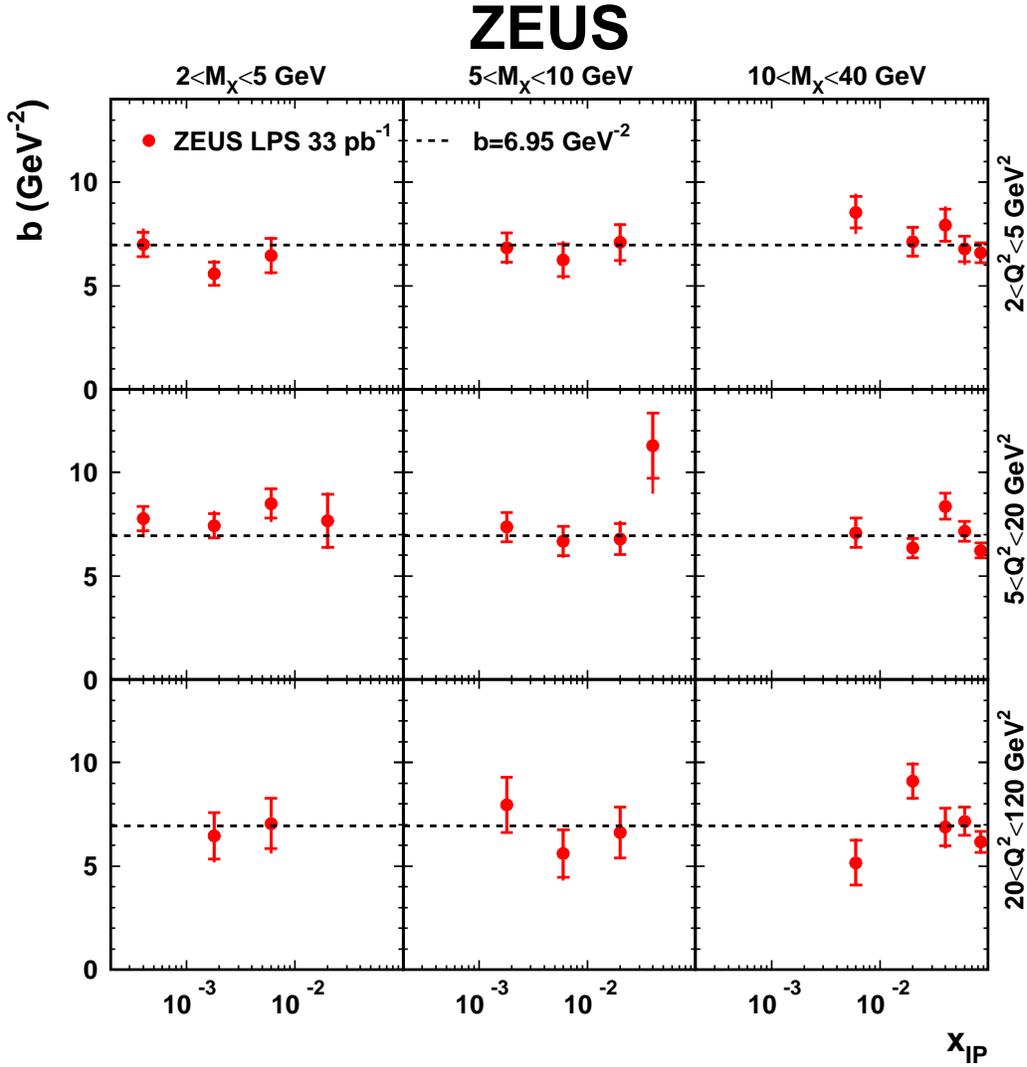} 
\end{center}
\caption{ 
The value of the exponential-slope parameter $b$ of the 
differential cross-section 
$d\sigma^{ep \rightarrow eXp}/dt \propto e^{-b|t|}$ as a function of 
$\xpom$ in 
bins of $Q^2$ and $M_X$. The dashed line corresponds to the value of 
$b = 6.95$~{\rm GeV}$^{-2}$, the average $t$-slope over the measured 
region. 
The inner error 
bars show the statistical uncertainties and the full bars indicate the 
statistical and the systematic uncertainties added in quadrature. 
}
\label{fig-tslopes} \vfill
\end{figure}

The values of the $t$-slope 
in different bins of $Q^2$, $M_X$ and $\xpom$ are shown in 
Fig.~\ref{fig-tslopes} and given in 
Table~\ref{tab-tslopes}. The diffractive-peak 
as well as the high-$\xpom$ region are shown. The $t$-slope does not 
depend on $Q^2$, $M_X$ and $\xpom$ in the measured regions. The lack of 
$Q^2$ dependence in a wide range of $Q^2$ as well as a value of $b$ much 
larger than that measured in hard diffraction (as discussed in a recent 
ZEUS publication~\cite{Chekanov:2007zr}) suggests that inclusive diffractive 
dissociation in DIS is a soft process.

\subsection{\boldmath{$\Phi$} dependence}
\label{sec-phi}

The azimuthal angle, $\Phi$, between the positron and proton scattering 
planes is sensitive to the helicity structure of the reaction $ep \to 
eXp$, as shown explicitly in Eq.~(\ref{fullsigma}).  To reduce the $\Phi$ 
dependence of the acceptance, an additional radial cut of 18\,cm was 
imposed on the impact point of the scattered positron at the RCAL 
surface, along with the restriction $Q^2>4$\,GeV$^2$. These cuts were only 
applied for the $\Phi$ analysis. The LPS data were used.

\begin{figure}[p] \vfill \begin{center}
\includegraphics[width=15cm,height=15cm]{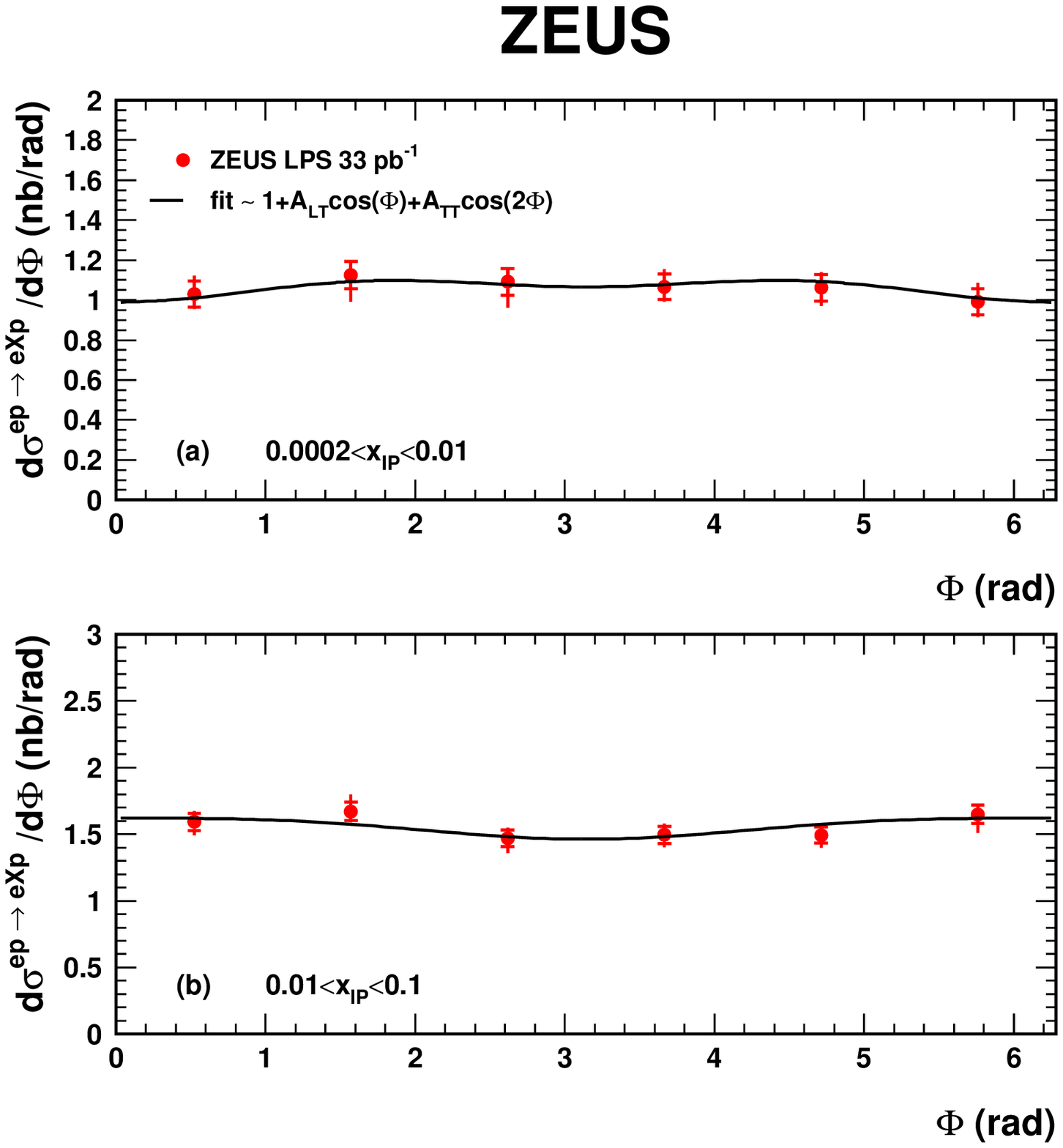} 
\end{center}
\caption{ 
The differential cross-section $d\sigma^{ep \rightarrow eXp}/d\Phi$ in the 
kinematic ranges (a) $0.0002<x_{\pom}<0.01$ and (b) $0.01<x_{\pom}<0.1$. 
The line shows the result of the fit described 
in~Section\protect\ref{sec-phi}. 
The inner error 
bars show the statistical uncertainties and the full bars indicate the 
statistical and the systematic uncertainties added in quadrature.
The normalisation uncertainty of $^{+11}_{-7}\%$ is not shown. 
} 
\label{fig-phi} \vfill
\end{figure}

The $\Phi$ distribution for the two ranges $0.0002<x_{\pom}<0.01$ and 
$0.01<x_{\pom}<0.1$ is presented in Figs.~\ref{fig-phi}a--b 
and Table~\ref{tab-phi}.

The distributions were fitted to the form
\begin{equation*}
\frac{d\sigma^{ep\rightarrow eXp}}{d\Phi}
\propto 1+A_{LT}\cos{\Phi}+A_{TT}\cos{2\Phi},
\end{equation*}
where $A_{LT}$ and $A_{TT}$ are proportional to
$\sigma^{\gamma^{\star}p \rightarrow Xp}_{LT}$ and
$\sigma^{\gamma^{\star}p \rightarrow Xp}_{TT}$, respectively.
The values of the azimuthal asymmetries are

\begin{eqnarray}
A_{LT} &=& -0.036 \pm 0.036 (\rm stat.) ^{+0.016}_{-0.014} (\rm
syst.),\nonumber\\
A_{TT}& =& -0.030 \pm 0.037 (\rm stat.) ^{+0.022}_{-0.006} (\rm 
syst.)\nonumber 
\end{eqnarray}
and
\begin{eqnarray} 
A_{LT} &=& +0.051 \pm 0.024 (\rm stat.)^{+0.012}_{-0.011} (\rm
syst.),\nonumber\\
A_{TT} &=&- 0.010 \pm 0.024 (\rm stat.)^{+0.010}_{-0.009} (\rm
syst.)\nonumber
\end{eqnarray}
for  the ranges $0.0002<x_{\pom}<0.01$ and $0.01<x_{\pom}<0.1$,
respectively.

The interference terms between the longitudinal and transverse amplitudes
and between the two transverse amplitudes are thus small in the
measured kinematic range, both in the diffractive-peak region and
at higher-$\xpom$ values, suggesting that the helicity structure of the
reaction $ep~\rightarrow~eXp$ is similar for both Pomeron and 
sub-leading Regge trajectories. 
   
\begin{figure}[p] \vfill \begin{center}
\includegraphics[width=15cm,height=15cm]{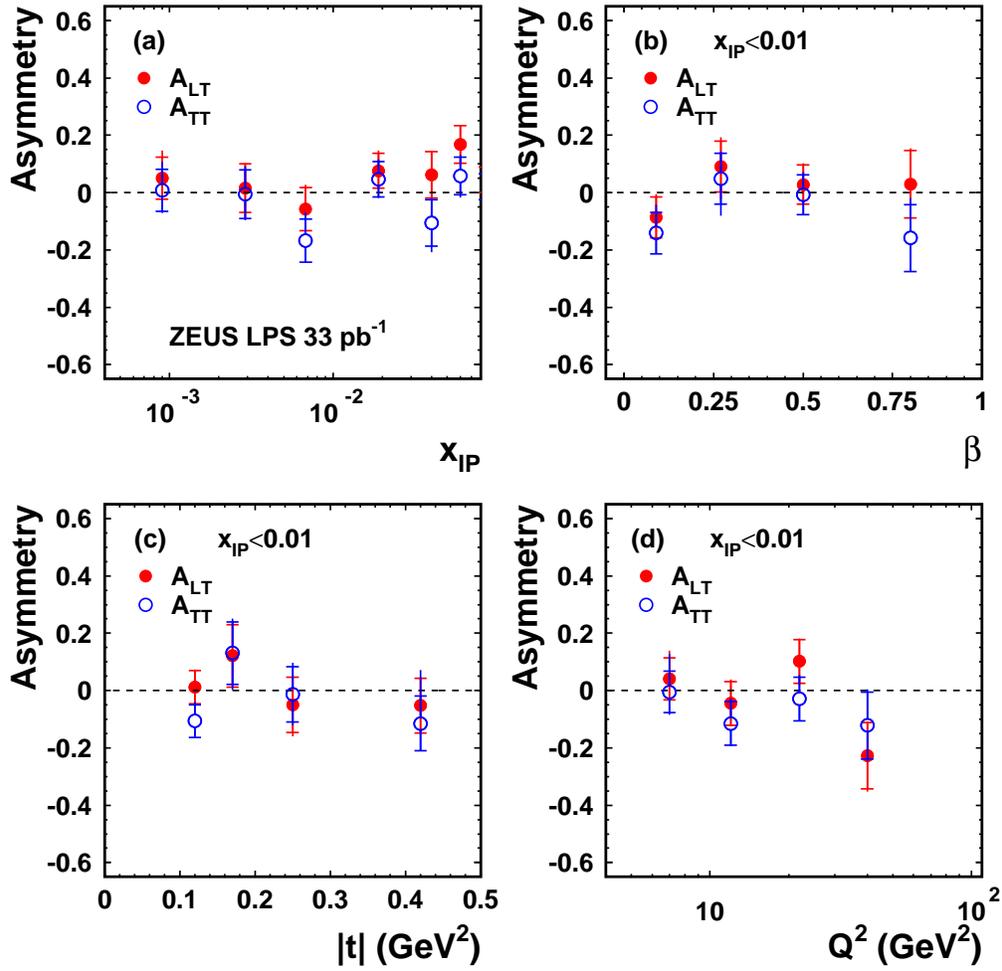} 
\end{center}
\caption{ 
The azimuthal asymmetries $A_{LT}$ (open circles) and $A_{TT}$ (dots) 
as a function of (a) $x_{\pom}$ and (b) $\beta$, (c) $|t|$ and (d) $Q^2$ 
for $\xpom<0.01$. 
The inner error 
bars show the statistical uncertainties and the full bars indicate the 
statistical and the systematic uncertainties added in quadrature.
} 
\label{fig-phi2} \vfill
\end{figure}

Figure~\ref{fig-phi2} presents $A_{LT}$ and $A_{TT}$ as a function of 
$x_{\pom}$, and, for $\xpom <0.01$, as a function of $\beta$, $t$ and 
$Q^2$. The asymmetries, also given in 
Tables~\ref{tab-phi-asymalt} and \ref{tab-phi-asymatt}, are consistent 
with zero.

The measured values of $A_{LT}$ can be compared with the results obtained 
in the exclusive electroproduction of $\rho^0$ mesons, $ep~\rightarrow~e 
\rho^0 p$, in which the hadronic final state, $X$, consists of a $\rho^0$ 
meson only. In this case, $A_{LT}=-\sqrt{2 \epsilon (1+\epsilon)} \cdot 
(r^5_{00}+2r^5_{11})= -0.256\pm 0.030 (\rm{stat.}) ^{+0.032}_{-0.022} 
(\rm{syst.})$, where $r^5_{00}$ and $r^5_{11}$ are two of the $\rho^0$ 
spin-density matrix elements~\cite{Chekanov:2007zr}. The present data 
therefore show that the asymmetry is smaller for inclusive scattering 
than for exclusive $\rho^0$ electroproduction.

There are numerous pQCD-based predictions for the behaviour of 
$A_{LT}$~\cite{gehrmann, arens, diehl, nikolaev} in the diffractive peak 
region, 
mostly for $\beta~\gsim~0.9$, where the asymmetry is expected to 
be largest, reflecting the dominance of 
$\sigma_L^{\gamma^{\star}p\rightarrow Xp}$ at large $\beta$ values. In 
all calculations, back-to-back configurations, i.e. $A_{LT}<0$, are 
favoured. There is no indication of such a behaviour in the present data; 
the statistics at high $\beta$ is however limited. The asymmetry is 
expected to be close to zero at low $\beta$, in agreement with the data.

\subsection{The reduced cross-section \boldmath{$\sigma_r^{D(4)}$} }
\label{sec-f2d4}

The LPS data are presented in Fig.~\ref{fig-f2d4_vs_xpom} in terms of the 
reduced cross-section $\sigma_r^{D(4)}$ in two $t$ bins,  
$0.09<|t|<0.19$\,GeV$^2$ and $0.19<|t|<0.55$\,GeV$^2$, with $\langle
|t| \rangle$~=~0.13\,GeV$^2$ and $\langle |t| \rangle$~=~0.3\,GeV$^2$,
respectively. The figure shows 
$x_{\pom}\sigma_r^{D(4)}$, also 
given in Tables~\ref{tab-xpsrd4lps-a} and \ref{tab-xpsrd4lps-b}, 
as a function of 
$x_{\pom}$ for different values of $\beta$, $Q^2$ and $|t|$.

At low $x_{\pom}$ and high $\beta$, 
$x_{\pom}\sigma_r^{D(4)}$ decreases with increasing $x_{\pom}$. 
At medium $\xpom$ and $\beta$, the dependence of $x_{\pom}\sigma_r^{D(4)}$ 
on $x_{\pom}$ is weak, whereas at high $\xpom$ and low $\beta$, 
$x_{\pom}\sigma_r^{D(4)}$ 
increases with increasing $x_{\pom}$.
The behaviour observed at high $\xpom$ and low $\beta$ can be ascribed to 
Reggeon and pion exchange. The Regge fit described in 
Section~\ref{intercept} indicates that the shape of the $\xpom$ 
dependence is the same in the two $t$ bins.

\subsection{The reduced cross-section \boldmath{$\sigma_r^{D(3)}$} }
\label{sec-f2d3}

\begin{figure}[p]
\vfill
\begin{center}
\includegraphics[width=15cm,height=15cm]{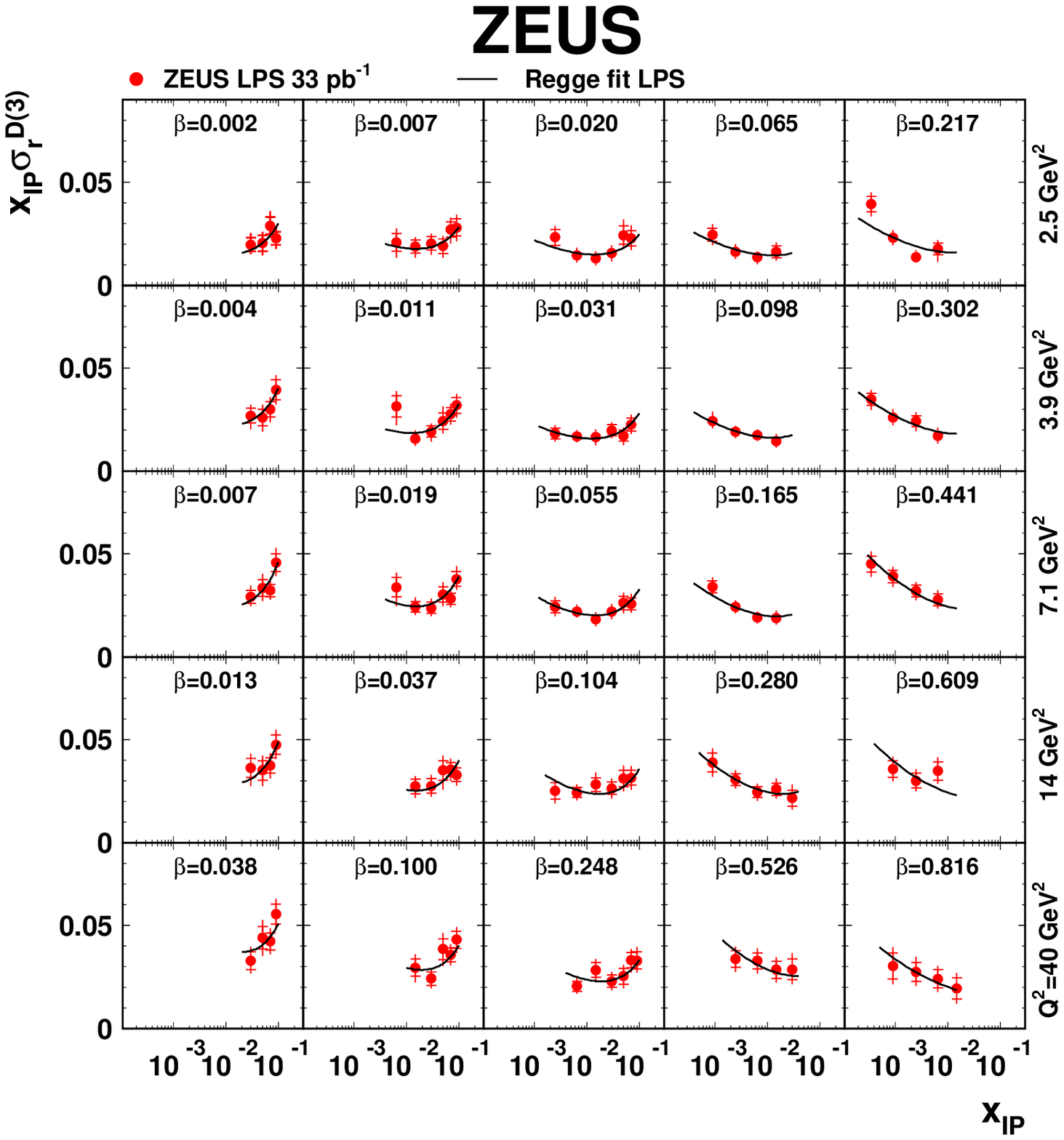}
\end{center}
\caption{
The reduced diffractive cross section multiplied by $\xpom$, $\xpom 
\sigma_r^{D(3)}$, obtained with the LPS method as
a function of $\xpom$ for different values of $Q^2$ and $\beta$. 
The lines are the result of the Regge fit described in 
Section~\protect\ref{intercept}.
The inner error 
bars show the statistical uncertainties and the full bars indicate the 
statistical and the systematic uncertainties added in quadrature.
The normalisation uncertainty of  $^{+11}_{-7}\%$ is not shown.
}
\label{fig-f2d3_vs_xpom}
\vfill
\end{figure}

\begin{figure}[p]
\vfill
\begin{center}
\includegraphics[width=15cm,height=15cm]{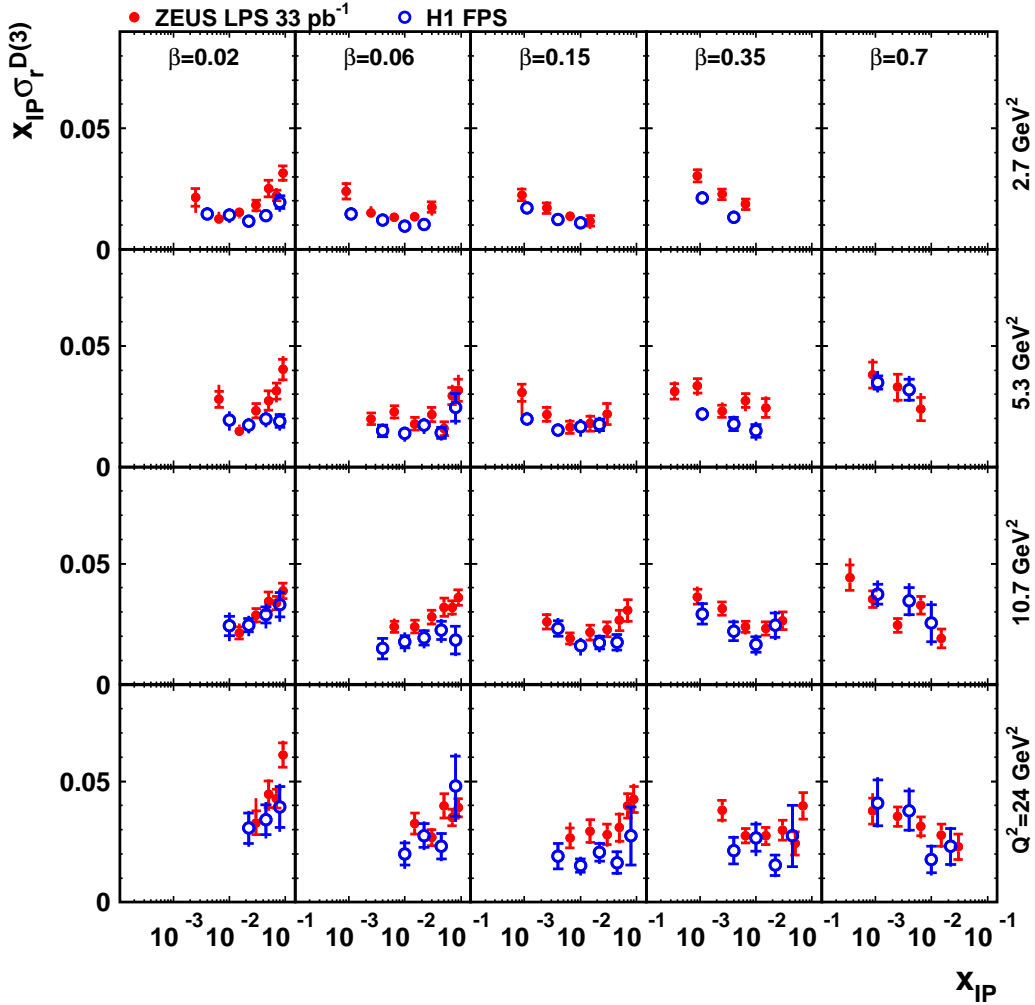}
\end{center}
\caption{
The reduced diffractive cross section multiplied by $\xpom$, $\xpom 
\sigma_r^{D(3)}$, obtained with the LPS method (dots) as a function of $\xpom$ 
for different values of $Q^2$ and $\beta$ compared with the results 
obtained with the H1 Forward Proton Spectrometer 
(open circles). 
The inner error 
bars show the statistical uncertainties and the full bars indicate the 
statistical and the systematic uncertainties added in quadrature.
The normalisation uncertainty of $^{+11}_{-7}\%$ of the ZEUS data is not 
shown, nor is that of the H1 data ($\pm 10\%$).  
}
\label{fig-lps-vs-fps}
\vfill
\end{figure}

The reduced cross section, $\xpom \sigma_r^{D(3)}$, obtained with the LPS 
method, is shown in Fig.~\ref{fig-f2d3_vs_xpom} and given in 
Table~\ref{tab-xpsrd3lps} as a function of $x_{\pom}$ for different values 
of $\beta$ and $Q^2$. The same features already discussed for $\xpom 
\sigma_r^{D(4)}$ are seen here. The LPS data are also shown in 
Fig.~\ref{fig-lps-vs-fps} compared with the H1 data from the H1 forward 
proton spectrometer (FPS)~\cite{h1-fps}. For this plot, the analysis was 
redone using the same $Q^2$ and $\beta$ bins as H1, thus avoiding 
extrapolation uncertainties.  The agreement is satisfactory.

\begin{figure}[p]
\vfill
\begin{center}
\includegraphics[width=15cm,height=15cm]{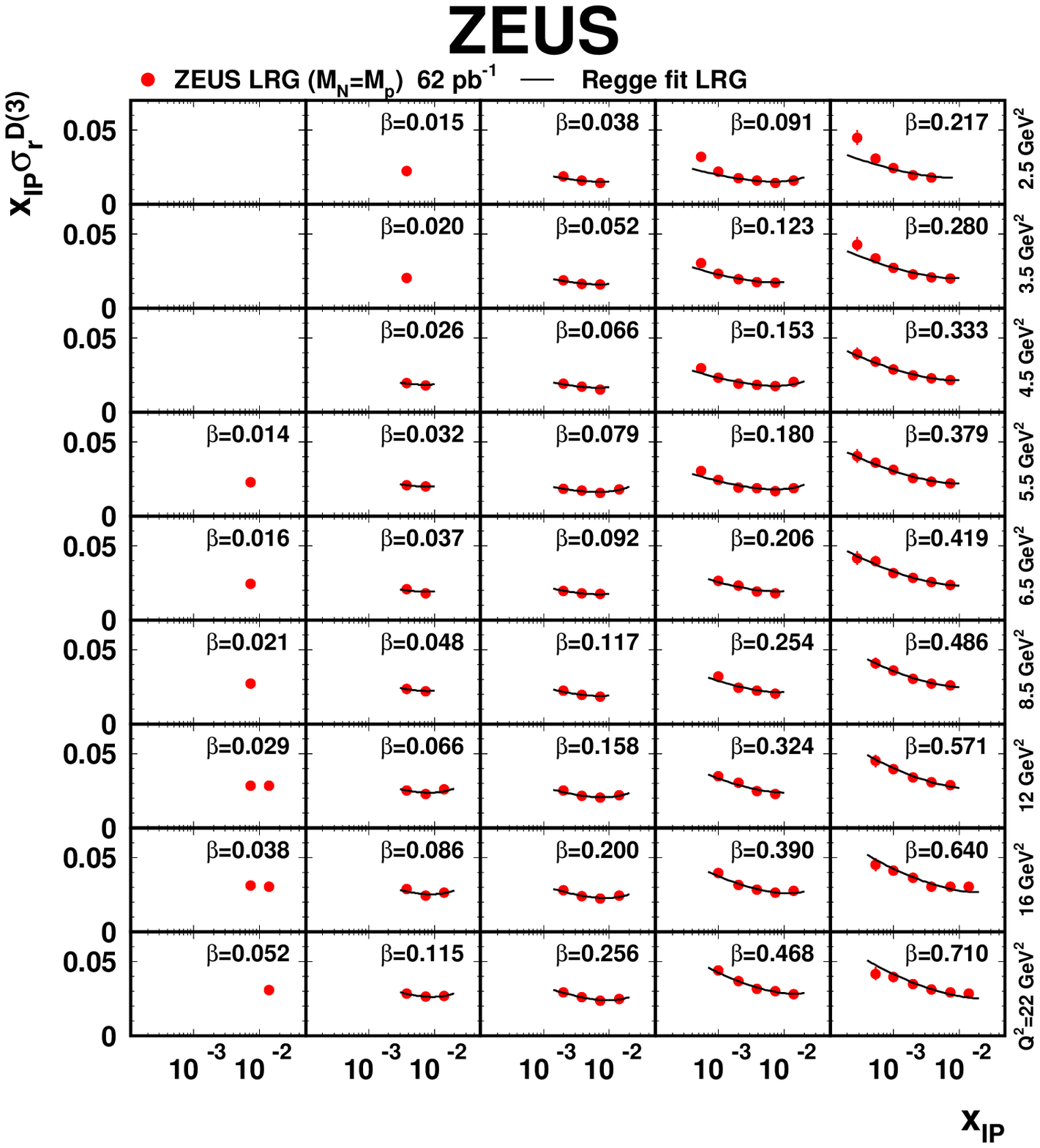}
\end{center}
\caption{
The reduced diffractive 
cross section multiplied by $\xpom$, $\xpom \sigma_r^{D(3)}$, 
obtained with the LRG method as a function of $\xpom$ for different 
values of $Q^2$ and $\beta$ at low $Q^2$ values. 
The lines are the result of the Regge fit described in 
Section~\protect\ref{intercept}.
The inner error 
bars show the statistical uncertainties and the full bars indicate the 
statistical and the systematic uncertainties added in quadrature.
The data are corrected for the proton-dissociative background to 
$M_N = M_p$ as described in Section~\protect\ref{sec-bkglrg}. 
The normalisation uncertainty of $\pm 5\%$ is not shown. 
}
\label{fig-f2d3-lrg_vs_xpom-a}
\vfill
\end{figure}

\begin{figure}[p] \vfill \begin{center} 
\includegraphics[width=15cm,height=15cm]{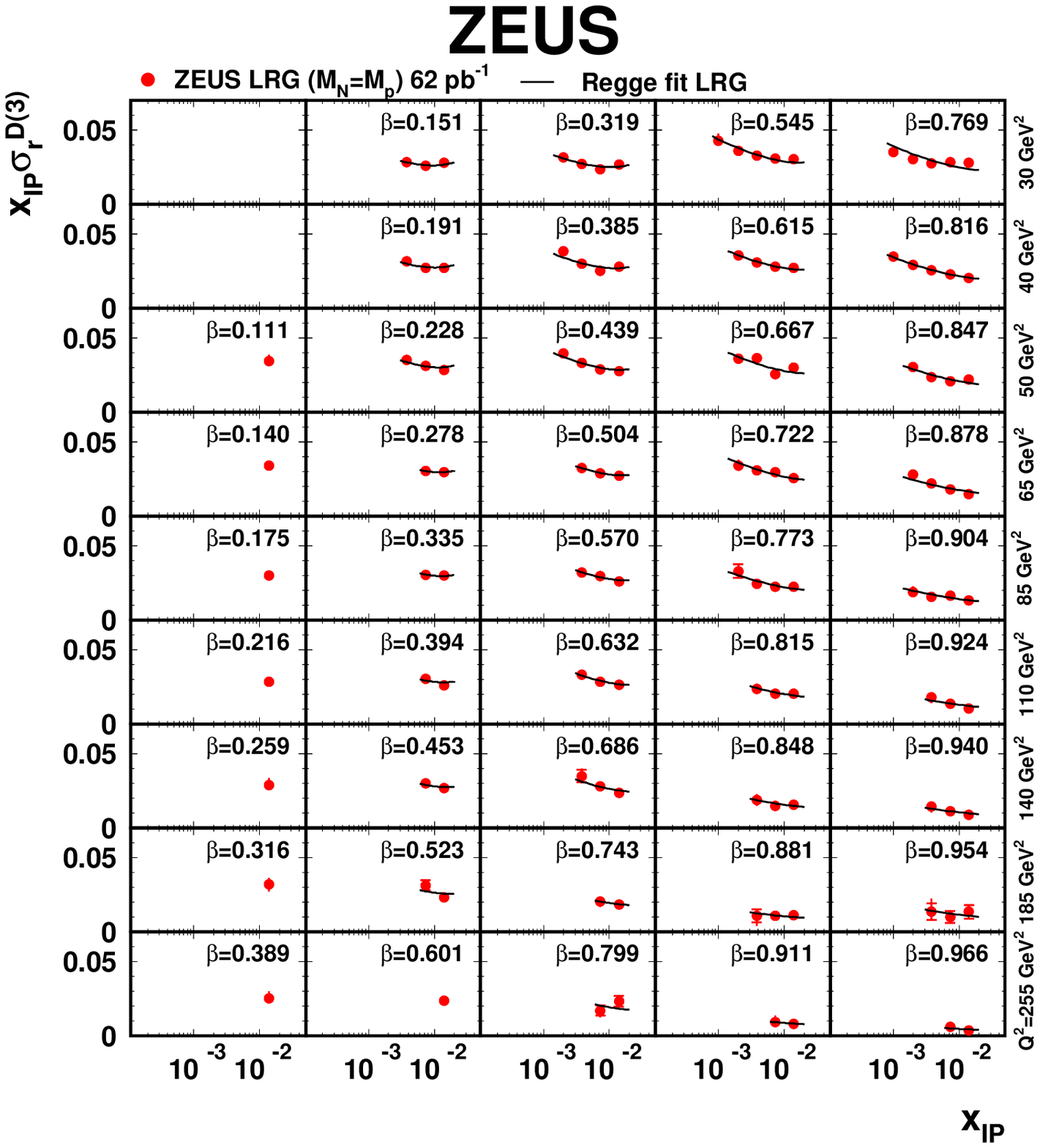} 
\end{center} 
\caption{ The reduced diffractive cross section multiplied by $\xpom$, 
$\xpom \sigma_r^{D(3)}$, obtained with the LRG method as a function of 
$\xpom$ for different values of $Q^2$ and $\beta$ at high $Q^2$ values. 
Other details as in caption for Fig.~\ref{fig-f2d3-lrg_vs_xpom-a}.
} 
\label{fig-f2d3-lrg_vs_xpom-b} 
\vfill \end{figure}

The LRG data, corrected to $M_N=M_p$ as discussed in 
Sect.~\ref{sec-bkglrg}, are presented in 
Figs.~\ref{fig-f2d3-lrg_vs_xpom-a} and \ref{fig-f2d3-lrg_vs_xpom-b} in 
terms of the reduced cross section, $\sigma_r^{D(3)}$. The figures show 
$x_{\pom}\sigma_r^{D(3)}$, also given in Table~\ref{tab-xpsrd3lrg}, as a 
function of $x_{\pom}$ for different values of $\beta$ and $Q^2$. The 
behaviour of $\xpom \sigma_r^{D(3)}$ is similar to that observed above for 
$\xpom \sigma_r^{D(4)}$, with an increase with decreasing $\xpom$ at low 
$\xpom$ and high $\beta$.

\begin{figure}[p]
\vfill
\begin{center}
\includegraphics[width=15cm,height=15cm]{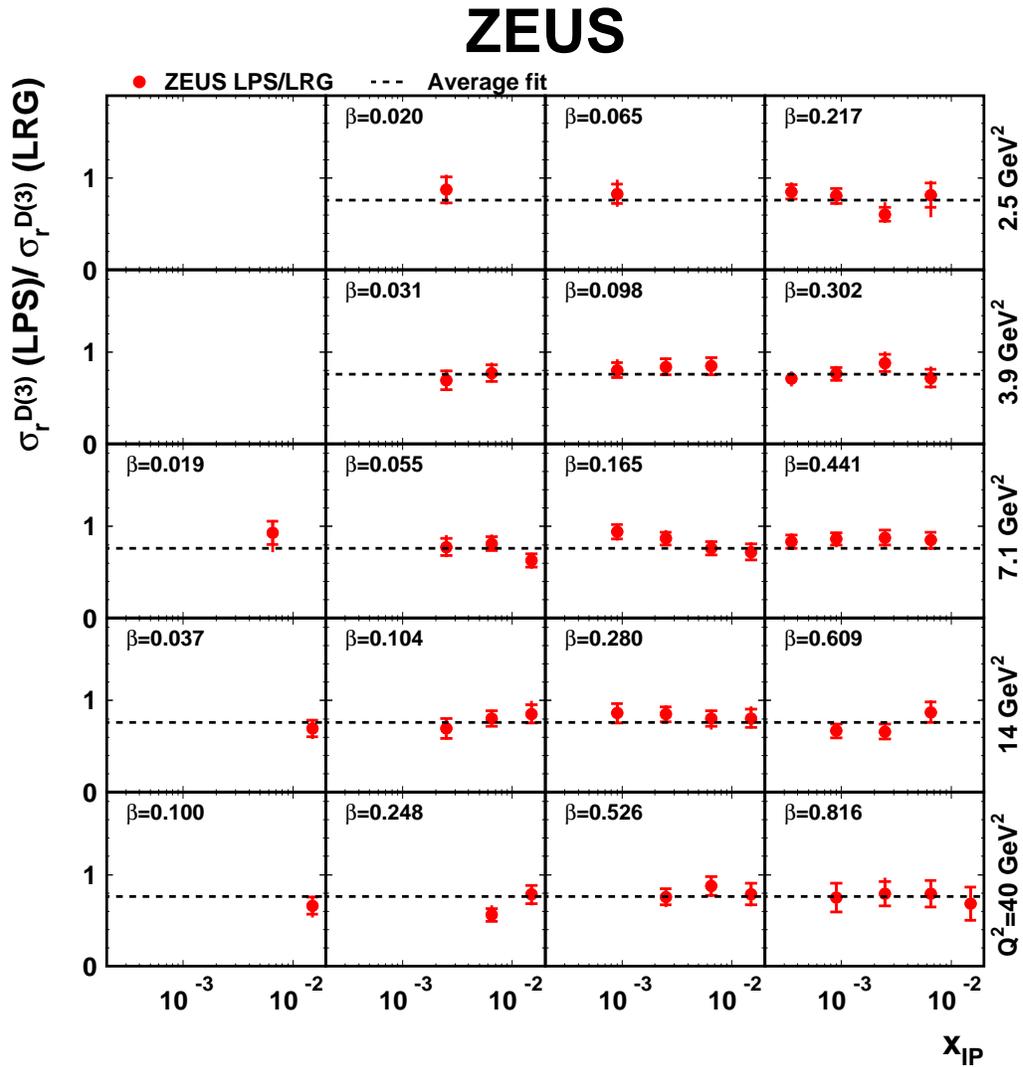}
\end{center}
\caption{
The ratio of the reduced diffractive cross sections, $\sigma_r^{D(3)}$, 
as 
obtained with the LPS and the LRG methods, before the subtraction of the 
proton-dissociative background, as a function of $\xpom$ for 
different values of $Q^2$ and $\beta$. 
The lines indicate the average value of the ratio. 
The inner error 
bars show the statistical uncertainties and the full bars indicate the 
statistical and the systematic uncertainties added in quadrature.
}
\label{fig-ratio_vs_xpom}
\vfill
\end{figure}

Figure \ref{fig-ratio_vs_xpom} shows the ratio of the $\sigma_r^{D(3)}$
values obtained with the LPS method to those obtained with the LRG 
method, before the subtraction of the proton-dissociative contribution.
The ratio is independent of $\xpom$, $Q^2$ and $\beta$ and averages
$0.76 \pm 0.01{\rm (stat.)}^{+0.03}_{-0.02}{\rm (syst.)}$. 
The $\xpom$, $Q^2$ and $\beta$ dependences of $\sigma_r^{D(3)}$ measured 
with the LPS method and
the LRG method are consistent in the region of overlap. The normalisation 
difference is ascribed to
the proton-dissociative contribution in the LRG sample, as discussed in 
Section~\ref{sec-bkglrg}. 

\begin{figure}[p]
\vfill
\begin{center}
\includegraphics[width=15cm,height=15cm]{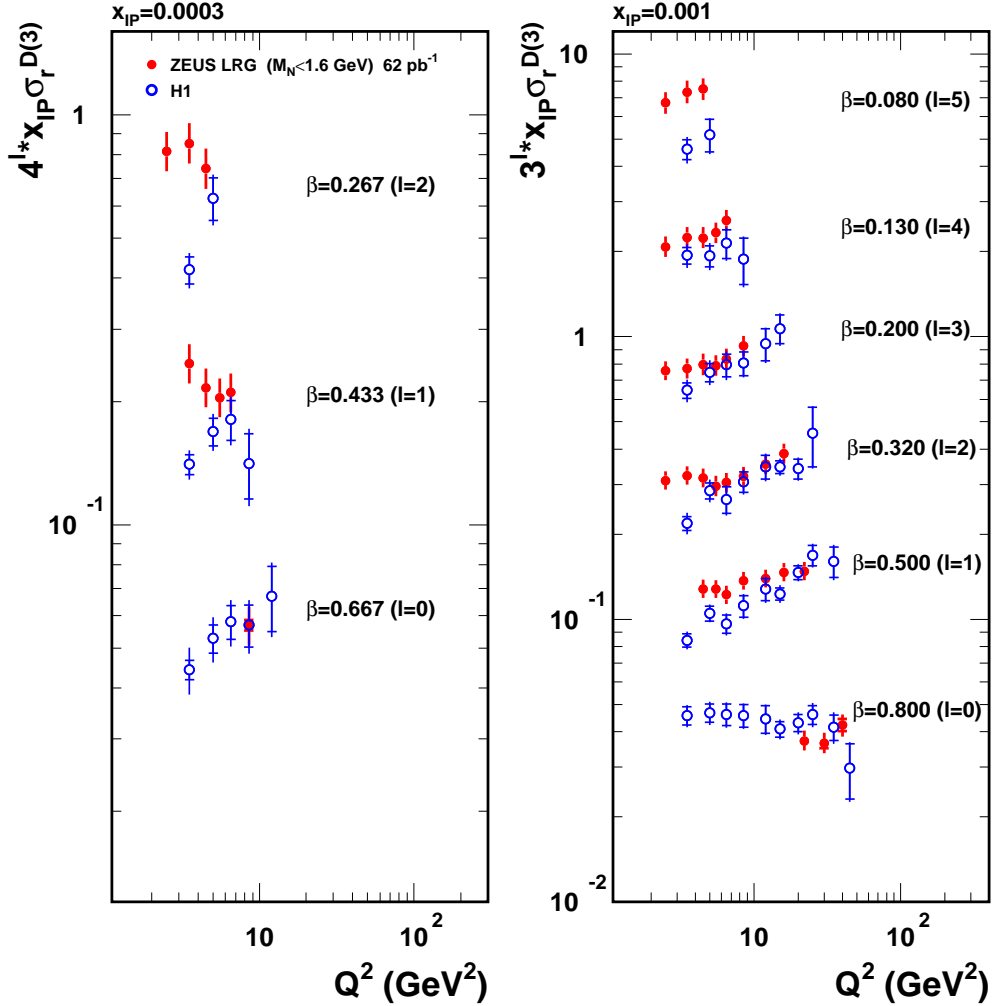}
\end{center}
\caption{
The reduced diffractive cross section multiplied by $\xpom$, $\xpom 
\sigma_r^{D(3)}$, obtained with the LRG method (dots) at $\xpom=0.0003$ 
and 
$\xpom=0.001$ as a function of  $Q^2$ for different $\beta$ values 
compared with the H1 results 
(open circles), also obtained with the LRG method. 
The inner error 
bars show the statistical uncertainties and the full bars indicate the 
statistical and the systematic uncertainties added in quadrature.
The ZEUS data are corrected to $M_N < 1.6$~{\rm GeV} as described in 
Section~\protect\ref{sec-f2d3}.
The $8\%$ uncertainty on the correction is not shown, nor is the 
$7\%$ relative normalisation uncertainty between the ZEUS and H1 data 
sets.
}
\label{fig-lrg-vs-h1a}
\vfill
\end{figure}

\begin{figure}[p]
\vfill
\begin{center}
\includegraphics[width=15cm,height=15cm]{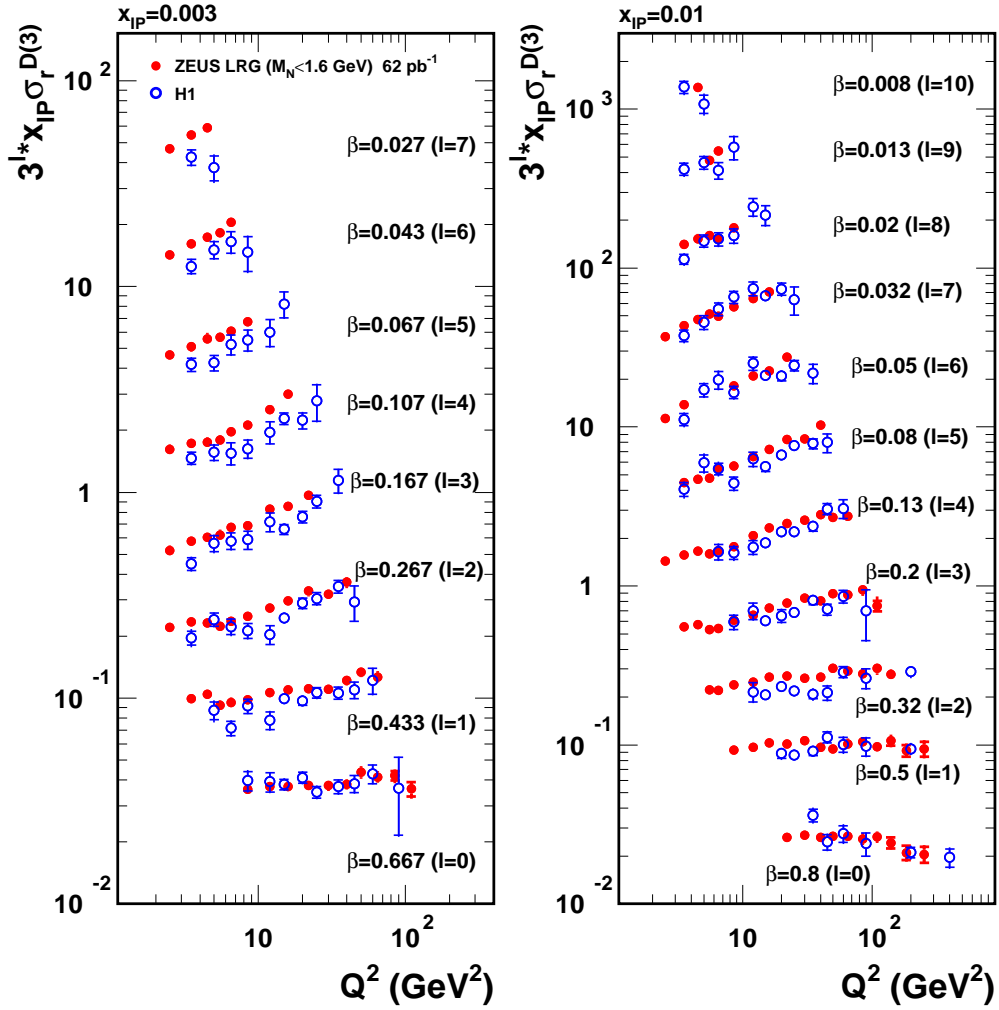}
\end{center}
\caption{
The reduced diffractive cross section multiplied by $\xpom$, $\xpom 
\sigma_r^{D(3)}$, obtained 
with the LRG method (dots) at $\xpom=0.003$ and $\xpom=0.01$ as a 
function of  $Q^2$ for different $\beta$ values
compared with the H1 results (open circles),  
also obtained with the LRG method. 
Other details as in caption for Fig.~\ref{fig-lrg-vs-h1a}.
}
\label{fig-lrg-vs-h1c}
\vfill
\end{figure}

The LRG data, corrected to $M_N < 1.6$\,GeV as described below, are shown 
as a function of $Q^2$ in different $\beta$ bins for $\xpom=0.0003$, 
$\xpom=0.001$, $\xpom=0.003$ and $\xpom=0.01$ in 
Figs.~\ref{fig-lrg-vs-h1a} and~\ref{fig-lrg-vs-h1c}. The values of $\xpom 
\sigma_r^{D(3)}$ exhibit a logarithmic rise with $Q^2$ for all $\beta$ 
values except in the lowest $\xpom$ bin ($\xpom=0.0003$) and in the 
highest $\beta$ bin ($\beta=0.8$). The rise observed even at high $\beta$
suggests that the diffractive PDFs of the proton are gluon-dominated.

In Figs.~\ref{fig-lrg-vs-h1a} and \ref{fig-lrg-vs-h1c} 
the LRG results are also compared with those of the H1 
Collaboration~\cite{h1-lrg}, similarly obtained with the LRG method. 
The ZEUS results are measured in the H1 $\beta$ and 
$\xpom$ bins; they are corrected to 
$M_N~<~1.6$\,GeV, as are the H1 data. 
The correction to $M_N<1.6$\,GeV for the present data, 
before the subtraction of the proton-dissociation background, 
was estimated with {\sc Pythia} to be 
$0.91 \pm 0.07$, independent of $\beta$, $Q^2$ and $\xpom$. 
Therefore, the ZEUS results in Figs.~\ref{fig-lrg-vs-h1a} 
and~\ref{fig-lrg-vs-h1c} were scaled down by $0.91$. 
With some exceptions, the shape agreement is reasonable. 
The ZEUS data are higher than the H1 data by $13\%$ on 
average, as estimated with a global fit to data for $Q^2 > 6$\,GeV$^2$. 
This normalisation discrepancy is 
consistent with the $8\%$ uncertainty on the proton-dissociation 
correction of $0.91 \pm 0.07$ combined with the $7\%$ relative 
normalisation uncertainty between the two data sets 
($\pm$7\% for H1 and $\pm$ 2.25\% for ZEUS). 

\begin{figure}[p]
\vfill
\begin{center}
\includegraphics[width=18cm]{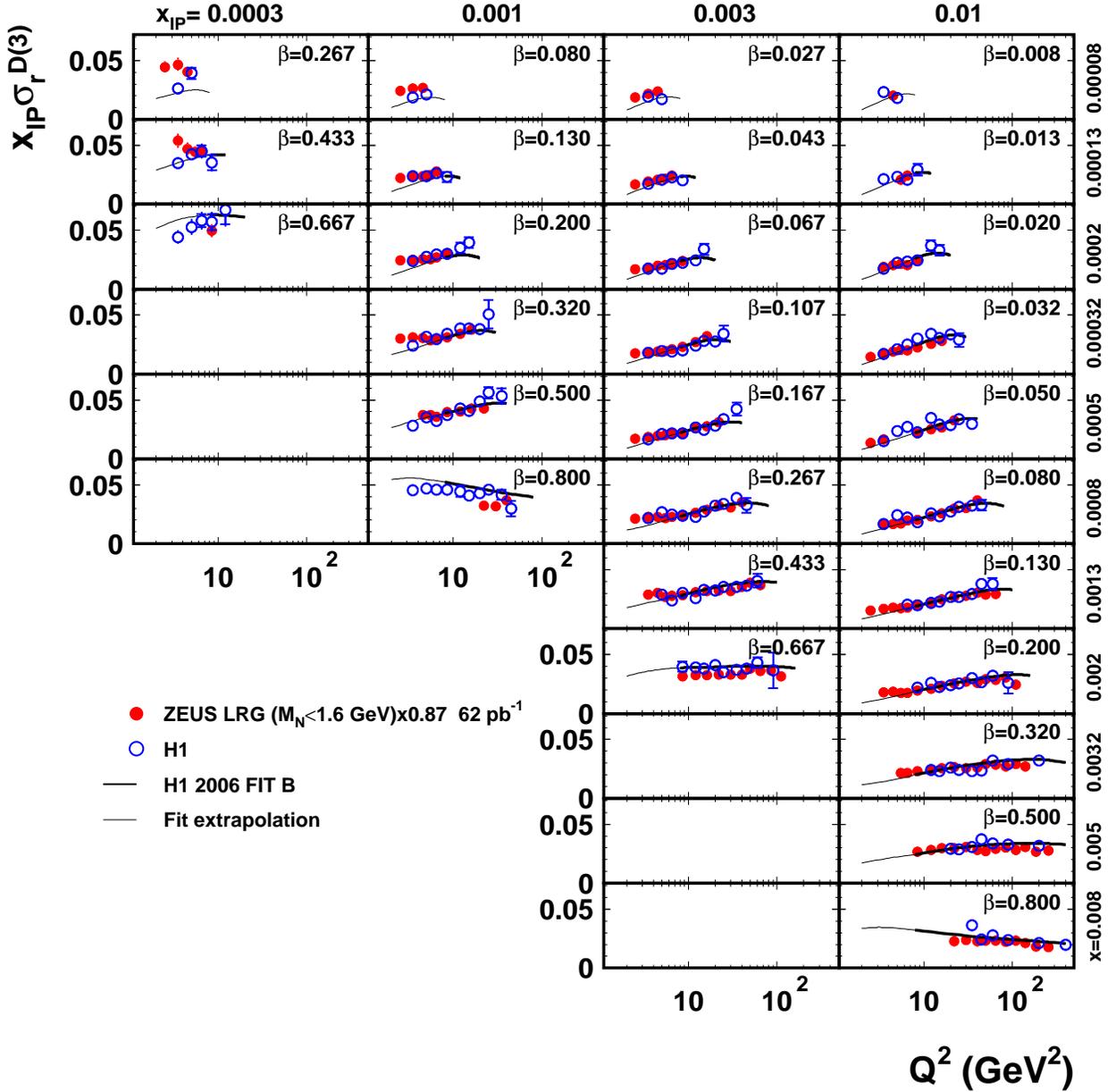}
\end{center}
\caption{
The reduced diffractive cross section multiplied by $\xpom$, $\xpom 
\sigma_r^{D(3)}$, obtained with the LRG method (dots) as a 
function of  $Q^2$ for different $\beta$ and $\xpom$ 
values compared with the H1 results (open circles), 
also obtained with the LRG method. The lines represent the 
expectation based on 
the diffractive parton distribution functions ``H1 2006 fit B''.
The inner error 
bars show the statistical uncertainties and the full bars indicate the 
statistical and the systematic uncertainties added in quadrature.
The ZEUS data are normalised to the H1 data. 
}
\label{fig-lrg-vs-h1-q2}
\vfill
\end{figure}

Figures~\ref{fig-lrg-vs-h1a} and \ref{fig-lrg-vs-h1c} are combined in 
Fig.~\ref{fig-lrg-vs-h1-q2} where the H1 and ZEUS reduced cross sections, 
the latter scaled down by the factor $1-0.13=0.87$ just described, are 
shown as 
a function of $Q^2$ in different $\beta$ and $\xpom$ bins. 
The result of the NLO QCD fit ``H1 2006 fit B''~\cite{h1-lrg} is
also shown. 
At fixed $\beta$, the $Q^2$ 
dependence of the two data sets, taken together, is different for 
different $\xpom$ values. Therefore, the data cannot be described by a 
single factorisable Regge contribution. 

\begin{figure}[p]
\vfill
\begin{center}
\includegraphics[width=15cm,height=15cm]{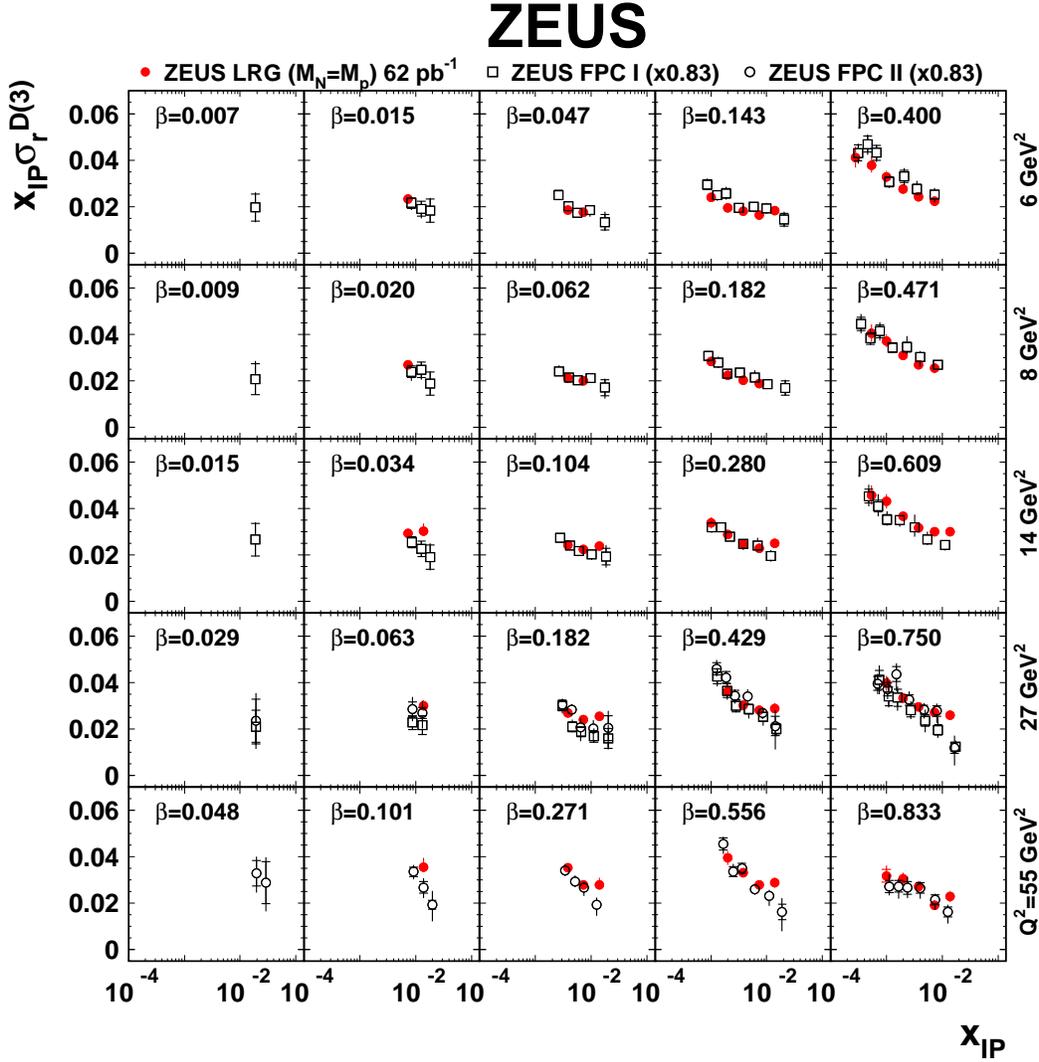}
\end{center}
\caption{
The reduced diffractive cross section multiplied by $\xpom$, $\xpom 
\sigma_r^{D(3)}$, obtained with the LRG method (dots) as a function of 
$\xpom$ for different values of $Q^2$ and $\beta$ at low $Q^2$, 
compared with the results obtained with the $M_X$ method, FPC~I (open squares) 
and FPC~II (open circles), 
scaled by the factor 0.83 described in Section~\protect\ref{sec-f2d3}.
The inner error 
bars show the statistical uncertainties and the full bars indicate the 
statistical and the systematic uncertainties added in quadrature.
}
\label{fig-lrg-vs-mxa}
\vfill
\end{figure}

\begin{figure}[p]
\vfill
\begin{center}
\includegraphics[width=15cm,height=15cm]{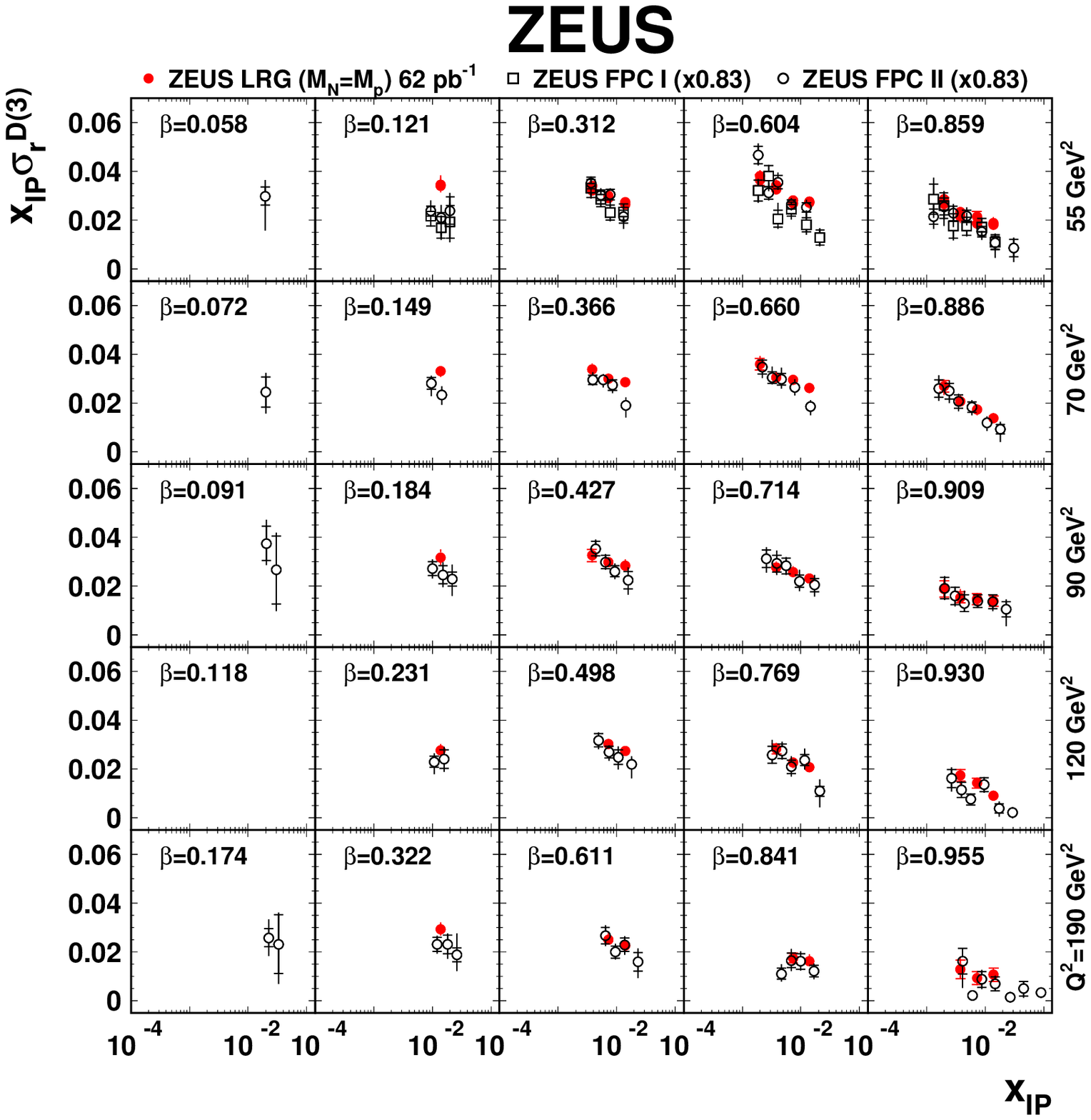}
\end{center}
\caption{
The reduced diffractive cross section multiplied by $\xpom$, $\xpom 
\sigma_r^{D(3)}$, obtained with the LRG method (dots) as a function of 
$\xpom$ for different values of $Q^2$ and $\beta$ at high $Q^2$.
Other details as in caption for Fig.~\ref{fig-lrg-vs-mxa}.
}
\label{fig-lrg-vs-mxb}
\vfill
\end{figure}

Figures \ref{fig-lrg-vs-mxa} and \ref{fig-lrg-vs-mxb} compare the LRG 
results, corrected to $M_N=M_p$, to those obtained with the $M_X$ 
method, referred to as FPC~I~\cite{mx1} and FPC~II~\cite{mx2}.
The LRG and FPC~II data were collected simultaneously; the 
two samples overlap by about $75\%$. The LRG results 
were recalculated in the bins used for the $M_X$-method results.  
The latter are 
for $M_N<2.3$\,GeV, but have been normalised here to the LRG results.
The scaling factor 
applied to the $M_X$ results was 0.83 $\pm$ 0.04, estimated 
with a global fit to the present data and the $M_X$ data; this factor 
quantifies the amount of residual
proton-dissociative background in the $M_X$ method. The overall agreement 
between the two measurements 
is reasonable. The different $\xpom$ dependence, more evident at low 
$Q^2$, may be ascribed to the fact that in the $M_X$ results the 
contribution of the Reggeon and pion trajectories is suppressed. 
In the low-$Q^2$ 
region, the $Q^2$ behaviour is somewhat different in the two data 
sets, with the $M_X$-method results decreasing faster with $Q^2$ than the 
LRG results.

\subsection{Extraction of the Pomeron trajectory}
\label{intercept}

In the framework of Regge phenomenology, the $\xpom$ dependence of 
$F_2^{D(4)}$ and $F_2^{D(3)}$ is related to the intercept of the Pomeron 
trajectory, the parameter that drives the energy dependence of the total 
hadron-hadron cross section at high energies~\cite{regge}. The Pomeron 
intercept in soft hadronic interactions is 
$1.096^{+0.012}_{-0.009}$~\cite{cudell}. However, the same parameter is 
significantly larger in the diffractive production of heavy vector mesons, 
notably in $J/\psi$ photoproduction (see e.g.~\cite{recent_review, 
Ivanov:2004ax}), reflecting the rapid rise of the cross section with $W$. 
This is a consequence of the increase of the parton densities in the 
proton at low $x$, which drives the rise of the cross section with 
decreasing $x$, and hence with decreasing $\xpom$ (since $\xpom \propto 
1/W^2 \propto x$).  The slope of the Pomeron trajectory, $\alpha'_{\pom}$, 
is smaller in the 
diffractive production of vector mesons~\cite{Ivanov:2004ax} than in 
soft hadron-hadron collisions, where 
$\alpha'_{\pom}=0.25$\,GeV$^{-2}$~\cite{rev}.  It is therefore interesting to 
determine if such deviations from the behaviour of the hadron-hadron data 
are also apparent in the inclusive diffractive dissociation of virtual photons. 

\begin{figure}[p]
\vfill
\begin{center}
\includegraphics[width=15cm,height=15cm]{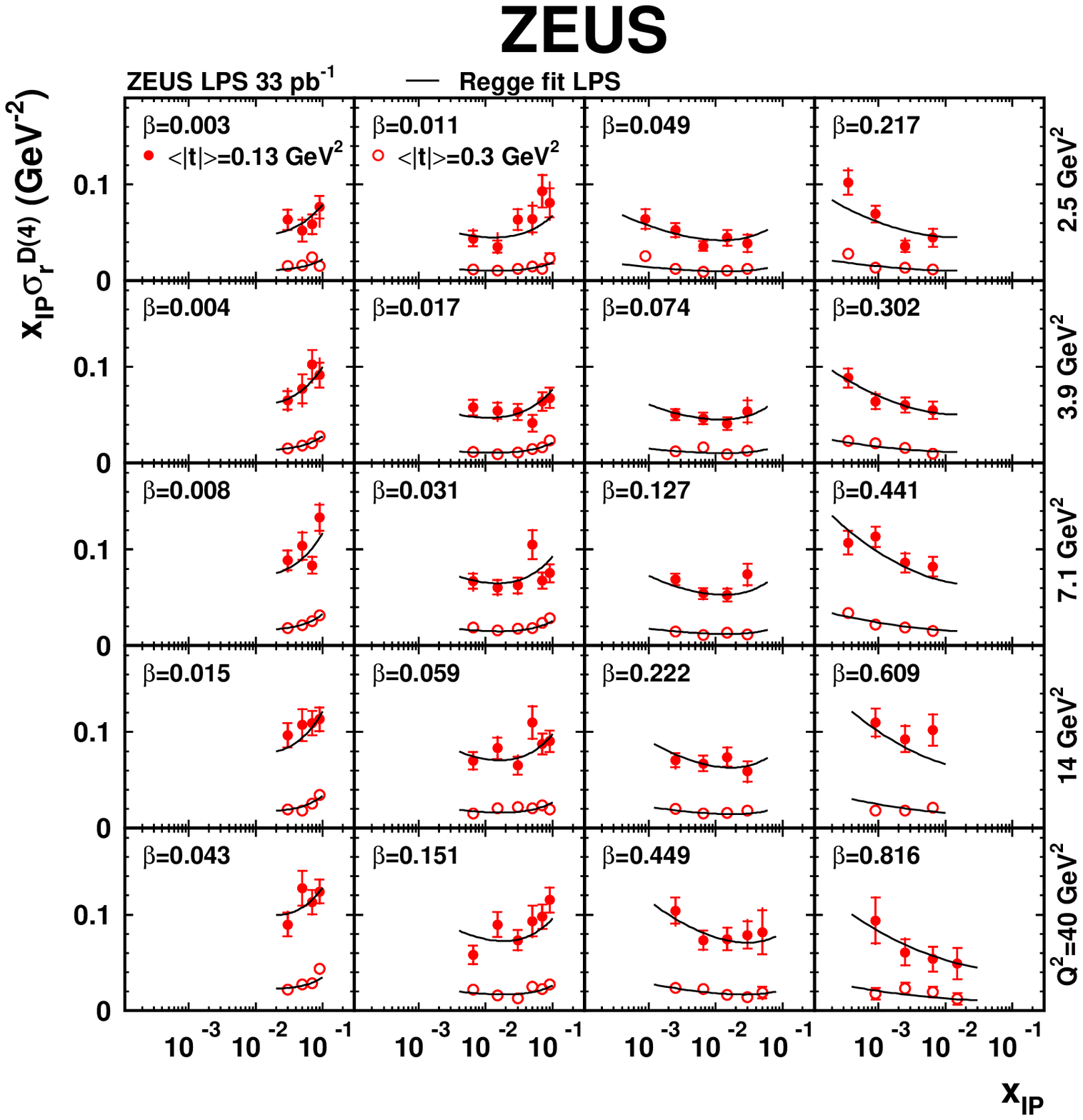}
\end{center}
\caption{
The reduced diffractive 
cross section multiplied by $\xpom$, $\xpom \sigma_r^{D(4)}$, 
obtained with the LPS method 
in two $t$ bins as a function of $\xpom$ for different values of $Q^2$ 
and $\beta$. The lines are the result of the Regge fit described in 
Section~\protect\ref{intercept}. The inner error 
bars show the statistical uncertainties and the full bars indicate the 
statistical and the systematic uncertainties added in quadrature.
The normalisation uncertainty of $\pm 7\%$ is not shown.  
}
\label{fig-f2d4_vs_xpom}
\vfill
\end{figure}

Following the Regge factorisation assumption (see Eq.~(\ref{i-s})), 
the data of Fig.~\ref{fig-f2d4_vs_xpom} were fitted to the form
\begin{equation*}
F_2^{D(4)}=f_{\pom}(x_{\pom},t)\cdot F_2^{\pom}(\beta,Q^2) + 
n_{\reg}\cdot f_{\reg}(x_{\pom},t)\cdot F_2^{\reg}(\beta,Q^2)~, 
\end{equation*}
where $n_{\reg}$ is a normalisation term. 
It was assumed that $F_2^{D(4)}= \sigma_r^{D(4)}$ and the fit 
was limited to $y<0.5$ to reduce the
influence of $F_L^D$. The Pomeron and the Reggeon fluxes were  
parameterised as~\cite{regge}
\begin{equation*}
f_{\pom}(x_{\pom},t)= 
\frac{e^{B_{\pom}t}}{x_{\pom}^{2\alpha_{\pom}(t)-1}}
\hspace{1cm}
{\rm and}
\hspace{1cm}
f_{\reg}(x_{\pom},t)=
\frac{e^{B_{\reg}t}}{x_{\pom}^{2\alpha_{\reg}(t)-1}}~,
\end{equation*}

\noindent and the Pomeron and Reggeon trajectories were both 
assumed to be linear.  
The fitted parameters were the Pomeron trajectory, $\alpha_{\pom}(0)$ and 
$\alpha'_{\pom}$, the intercept of the
Reggeon trajectory, $\alpha_{\reg}(0)$, 
the slope, $B_{\pom}$, 
and the Reggeon normalisation
term, $n_{\reg}$. The Reggeon structure function, $F_2^{\reg}(\beta,Q^2)$, 
was taken to be equal to 
the pion structure function as parameterised by
GRV~\cite{grv1,grv2,grv3}. The slope $B_{\reg}$ was fixed to 
$2.0$\,GeV$^{-2}$, taken from hadron-hadron data, and the slope of the 
Reggeon trajectory, $\alpha'_{\reg}$, was fixed to $0.9$\,GeV$^{-2}$. 
The lines in Fig.~\ref{fig-f2d4_vs_xpom} show the result of the fit. 
The results for the fit parameters are given in Table~\ref{tab-reggefitlps}. 

The model uncertainty reflects the effect of $R^D$, which was varied between
0 and 1, and that of the parameterisation of the pion structure function, 
which was changed from that of GRV to that of Owens \cite{pion}.
The quality of the fit is good. The Pomeron intercept 
is consistent with that of the soft Pomeron. The result for 
$\alpha'_{\pom}$ is significantly lower than
$\alpha'_{\pom}=0.25$\,GeV$^{-2}$; it agrees with the result 
recently found 
by the H1 Collaboration~\cite{h1-lrg} as well as with the values found 
in the diffractive production of vector mesons \cite{Ivanov:2004ax}.
The Reggeon intercept is higher than the expectation of 0.5475
based on the Donnachie and Landshoff fits to the $pp$, $\bar{p}p$, $Kp$,
$\pi p$ and $\gamma p$ total cross section data~\cite{softpom}. 
Allowing for maximal interference between the Pomeron and Reggeon 
amplitudes also gives a good fit.

A similar fit was performed to the $\sigma_r^{D(3)}$ LRG points. 
The result of the fit is shown in 
Figs.~\ref{fig-f2d3-lrg_vs_xpom-a} and \ref{fig-f2d3-lrg_vs_xpom-b} and the 
parameters, both those kept fixed and those obtained from the fit, 
are summarised in 
Table~\ref{tab-reggefitlrg}.
The first uncertainty is that from 
the fit, in which the quadratic sum of statistical and systematic 
uncertainties was used.  
The model uncertainty reflects the variation of $\alpha'_{\pom}$ between 0 
and 0.1\,GeV$^{-2}$ and that of $\alpha_{\reg}(0)$ between 0.55 and 0.75; 
in addition, as for the fit to the LPS data, 
$R^D$ was varied between 0 and 1, and the pion 
structure function parameterisation was changed from that of GRV to that 
of Owens \cite{pion}. Here again, the fit was limited to $y<0.5$. 
 The quality of the fit is very good.

\begin{figure}[p]
\vfill
\begin{center}
\includegraphics[width=15cm,height=15cm]{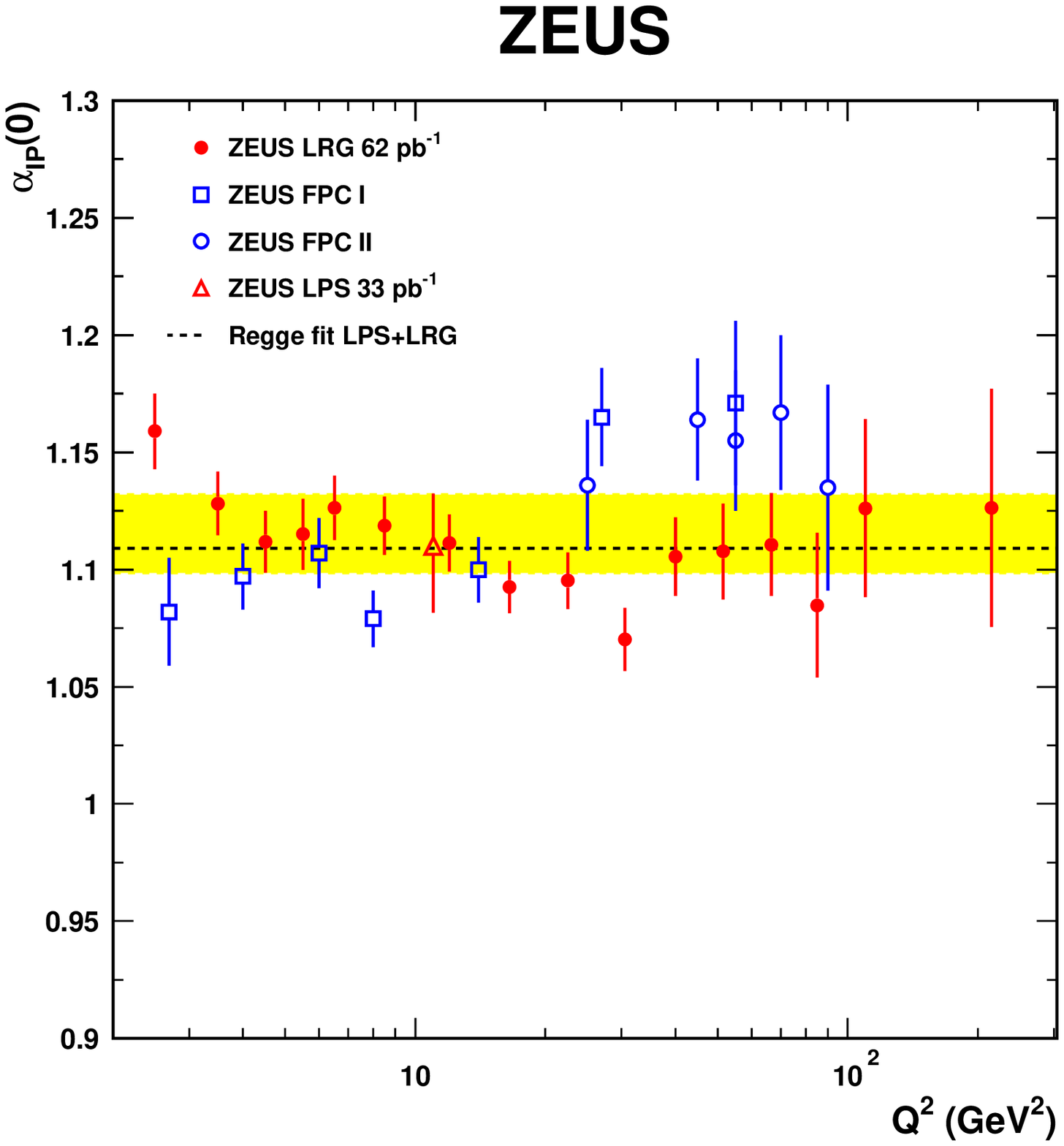}
\end{center}
\caption{
The Pomeron intercept $\alpha_{\pom}(0)$ as a function of $Q^2$ as 
obtained from the LRG (dots) and the LPS data (triangles). Also 
shown are the $M_X$-method results, FPC~I (open squares) and 
FPC~II (open circles). The error bars indicate the uncertainty from the 
fit for the LRG and FPC points; they indicate the statistical and 
systematic uncertainties summed in quadrature for the LPS points. 
The dashed line indicates the results of the 
Regge fit to the LPS and LRG data together described in 
Section~\protect\ref{intercept}, and the 
band indicates the size of the total error. 
}
\label{fig-alpha-vs-q2}
\vfill
\end{figure}

Figure \ref{fig-alpha-vs-q2} shows $\alpha_{\pom}(0)$ as a function of 
$Q^2$; it was obtained with a fit to the LRG data 
in bins of $Q^2$, similar to that 
described earlier for the full $Q^2$ range. The Reggeon normalisation 
term, $n_{\reg}$, was fixed to 
the value $n_{\reg} = 2.6 \pm 0.3$, extracted from a combined Regge fit 
to the LPS and LRG results in the full $Q^2$ range. The LPS result and 
those obtained with 
the $M_X$ method, FPC~I~\cite{mx1} and FPC~II~\cite{mx2}, are also shown. 
In the region explored, the present 
data do not exhibit a significant dependence on $Q^2$. The agreement  
with the $M_X$-method results is fair.

\section{Summary}
\label{sec-con}

Measurements have been presented of the reaction $ep~\rightarrow~eXp$ 
obtained by requiring a large rapidity gap in the forward direction (LRG 
sample) or the detection of a proton in the leading proton spectrometer 
(LPS sample). The kinematic region is $2<Q^2<305$\,GeV$^2$ (LRG) or 
$2<Q^2<120$\,GeV$^2$ (LPS), $40<W<240$\,GeV, $2<M_X<25$\,GeV 
(LRG) or $2<M_X<40$\,GeV (LPS), $0.0002<\xpom<0.02$ (LRG) 
or $0.0002<\xpom<0.1$ (LPS) and $0.09<|t|<0.55$\,GeV$^2$ (LPS).

The LPS data are presented in terms of the $t$ 
and $\Phi$ dependences of the cross section, as well as of the 
$\xpom$, $Q^2$, $\beta$ and $t$ dependences of the reduced 
diffractive cross section, $\sigma_r^{D(4)}$.
The $t$ dependence of the cross section is approximately 
exponential, with a $t$-slope $b=7.0 \pm0.3$\,GeV$^{-2}$. The 
slope is independent of $Q^2$, $M_X$ and $\xpom$. The lack of 
$Q^2$ dependence and the value of $b$ much larger than that measured in 
hard diffraction suggest that this is a soft process. 
There is no significant $\Phi$ dependence of the cross section.
The cross-section $\sigma_r^{D(4)}$
was measured for the first time in two $t$ bins and 
was found to have the same $\xpom$ dependence in the two bins. 

The reduced cross-section $\sigma_r^{D(3)}$ was measured using 
both the LRG and LPS data. Consistent results were found for the shape. 
The normalisation difference of about 25\% is ascribed to the 
proton-dissociative contribution in the LRG data.
Within the normalisation uncertainties the results agree reasonably 
well with the H1 measurements~\cite{h1-lrg}. The comparison with the ZEUS 
$M_X$-method results \cite{mx1, mx2} indicates that the latter have a 
residual 
proton-dissociative contribution of 17\%; the shape agreement is 
good, especially at low $\xpom$.
A Regge fit to $\sigma_r^{D(3)}$ supports the $Q^2$ independence of 
$\alpha_{\pom}(0)$.

\section*{Acknowledgements} \label{sec-ack} We thank the DESY Directorate
for their support and encouragement. We are grateful for the support of the
DESY computing and network services. We are specially grateful to the HERA
machine group: collaboration with them was crucial to the successful
installation and operation of the leading proton spectrometer.  The design,
construction and installation of the ZEUS detector have been made possible
by the ingenuity and effort of many people who are not listed as authors. 
It is a pleasure to thank A.D.~Martin, M.G.~Ryskin and G.~Watt for many 
useful discussions.

\vfill\eject

\providecommand{\etal}{et al.\xspace}
\providecommand{\coll}{Coll.\xspace}
\catcode`\@=11
\def\@bibitem#1{%
\ifmc@bstsupport
  \mc@iftail{#1}%
    {;\newline\ignorespaces}%
    {\ifmc@first\else.\fi\orig@bibitem{#1}}
  \mc@firstfalse
\else
  \mc@iftail{#1}%
    {\ignorespaces}%
    {\orig@bibitem{#1}}%
\fi}%
\catcode`\@=12

\clearpage

\caption{The values and uncertainties of the parameters extracted from the 
Regge fits to the LPS data, without interference
and with maximal interference between the Pomeron and Reggeon exchanges.
}
\label{tab-reggefitlps}
\end{center}
\end{table}

\vspace{2cm}

\begin{table}[h]
\begin{center}
\begin{tabular}{|c|c|}
\hline
 & Fit LRG\\
\hline
$\chi^2/{\rm ndf}$ & $\chi^2/{\rm ndf}=143.9/168$\\
\hline
$\alpha_{\pom}(0)$ & $1.117 \pm 0.006 ^{+0.022}_{-0.007} ({\rm model})$\\
\hline
$\alpha_{\reg}(0)$ & fixed to 0.75\\
\hline
$\alpha'_{\pom}  $ & fixed to $0$\\
\hline
$B_{\pom}$ & fixed to 7.0\,GeV$^{-2}$ \\
\hline
$n_{\reg}$ & $ 2.98 \pm 0.39$ \\
\hline
\end{tabular}
\caption{The values of the parameters extracted from the Regge fit to the 
LRG data and the corresponding uncertainties.}
\label{tab-reggefitlrg}
\end{center}
\end{table}


\clearpage

\clearpage

%
%
\end{document}